\documentclass[preprint2]{aastex}

\usepackage{booktabs}
\usepackage{lscape}

\newcommand{\vul}{Vul$\,$OB1}
\newcommand{\hii}{H\footnotesize II \normalsize }
\newcommand{\hi}{H\footnotesize I \normalsize }
\newcommand{\degmath}{\,^\circ}
\newcommand{\mic}{$\,\mu$m}
\newcommand{\mm}{$\,$mm}
\slugcomment{Draft}

\shorttitle{YSO and triggered star formation in \vul}
\shortauthors{Billot et al.}

\begin{document}

\title{Young Stellar Objects and Triggered Star Formation in the Vulpecula OB association}

\author{N. Billot\altaffilmark{1}, A. Noriega-Crespo\altaffilmark{2}, S. Carey\altaffilmark{2}, S. Guieu\altaffilmark{2},  S. Shenoy\altaffilmark{2}, R. Paladini\altaffilmark{2}, W. Latter\altaffilmark{1}}

\email{nbillot@ipac.caltech.edu}

\altaffiltext{1}{NASA Herschel Science Center, IPAC, MS 100-22, California Institute of Technology, Pasadena, CA 91125 USA}
\altaffiltext{2}{Spitzer Science Center, IPAC, MS 220-6, California Institute of Technology, Pasadena, CA 91125 USA}

\begin{abstract}

The Vulpecula OB association, \vul, is a region of active star formation located in the Galactic plane at 2.3$\;$kpc from the Sun. Previous studies suggest that sequential star formation is propagating along this 100$\;$pc long molecular complex. In this paper, we use \emph{Spitzer} MIPSGAL and GLIMPSE data to reconstruct the star formation history of \vul, and search for signatures of past triggering events. We make a census of Young Stellar Objects (YSO) in \vul~based on IR color and magnitude criteria, and we rely on the properties and nature of these YSOs to trace recent episodes of massive star formation. We find 856 YSO candidates, and show that the evolutionary stage of the YSO population in \vul~is rather homogeneous - ruling out the scenario of propagating star formation. We estimate the current star formation efficiency to be $\sim8$\%. We also report the discovery of a dozen pillar-like structures, which are confirmed to be sites of small scale triggered star formation.

\end{abstract}


\section{Introduction}

The physical mechanisms describing stellar birth are fairly well understood for low- and intermediate-mass stars but still under debate for their high-mass analogues \citep{mckee07}. 
On larger scales, observations show that star forming mechanisms are of relatively poor efficiency, as only a fraction of the gas reservoir in the Universe is turning into stars. Typical Star Formation Efficiencies (SFE) are of the order 3-6\% in the Galaxy \citep{evans09}, and 5\% or less in other galaxies \citep{rownd99}. Still, in extreme environments like in the starburst galaxy Arp~220, the SFE can reach 50\% \citep{anan00} suggesting that the star forming mechanism at work in such objects could be of a different nature. 
For instance the feedback into the interstellar medium (ISM) from short-lived massive stars seems to influence the yield of star formation in their local environment. As they evolve off the main sequence, high-mass stars produce a copious amount of energy while still embedded in their native cocoon; such disruption of a molecular cloud leads to gravitational instabilities and possibly to the onset of a new episode of star formation. \citet{hosokawa06}  showed that under certain conditions runaway triggering can take place around massive OB stars. In other cases however, turbulence and magnetic fields can have a negative feedback on the local ISM and lead to the suppression of the star forming activity \citep{price09,stone70}. 
\citet{elmegreen98} provides an exhaustive review of the theoretical framework to study the dynamical triggers of star formation.

In this context, we use the data from the \emph{Spitzer} Legacy surveys MIPSGAL \citep{carey09} and GLIMPSE \citep{benjamin03} to study the star forming activity currently taking place in the Vulpecula OB association (hereafter \vul).  \vul~hosts nearly one hundred OB~stars and three bright \hii regions known as Sharpless objects 86, 87 and 88 \citep{sharpless59}. According to \citet{ehlerova01} and \citet{turner86}, the star forming activity occurring in \vul~might have been triggered by a common external source, and star formation might be propagating from one \hii region to another through the expansion of a supernova shock front. The aim of this paper is to study the triggered star formation on scales as large as $\sim$100~pc in \vul, which could help us understand star formation mechanisms within other Giant Molecular Clouds in the Galaxy and beyond.

We use the method developed by \citet{gutermuth08} to obtain a census of Young Stellar Objects (hereafter YSOs) in \vul~based on infrared color and magnitude criteria, and we rely on the properties and nature of these YSOs to trace episodes of star formation. The MIPSGAL and GLIMPSE sensitivity makes our search for YSOs in \vul~biased towards massive objects that reveal the most recent star forming activity. 

In Section~\ref{sec:obs_cat} we present the dataset used in the paper. In Section~\ref{sec:vulOB1} we give a comprehensive description of \vul~encompassing the three \hii regions as well as the dozen pillar-like structures we discovered in this region. Section~\ref{sec:YSO} describes the identification process of Young Stellar Objects based on their IR-excess emission. We present and discuss our results in Section~\ref{sec:result}, and we give our conclusions in Section~\ref{sec:conclu}.

\section{The dataset}
\label{sec:obs_cat}

\subsection{\emph{Spitzer} Observations and Point Source Catalogs}
\label{subsec:obs_cat}

The \emph{Spitzer Space Telescope} \citep{werner04} observed over 270~square degrees of the inner Galactic plane in six wavelength bands as part of two legacy programs: the Galactic Legacy Infrared Mid-Plane Survey Extraordinaire \citep[GLIMPSE; E.~Churchwell PI;][]{benjamin03} and the MIPS GALactic plane survey \citep[MIPSGAL; S.~Carey PI;][]{carey09}. The Vulpecula region was covered by GLIMPSE$\,$I (PID~188) and MIPSGAL$\,$I (PID~20597). It was imaged with the IRAC camera \citep{fazio04} at 3.6, 4.5, 5.8 and 8.0\mic, and with the MIPS camera \citep{rieke04} at 24 and 70\mic. The angular resolution of \emph{Spitzer} is 2\arcsec, 6\arcsec~and 18\arcsec~at 8, 24 and 70\mic, respectively.

The MOPEX package \citep{makovoz06} was used to build mosaics of the sky about 1~square degree wide from individual Basic Calibrated Data (BCD) frames. Details of the post-processing and BCD pipeline modifications are described in \citet{meade07} for GLIMPSE and in \citet{mizuno08} for MIPSGAL. The large Vulpecula mosaics presented in this paper are also built with MOPEX using the 1~square degree plates available through the Infrared Science Archive\footnote{http://irsa.ipac.caltech.edu/data/SPITZER/GLIMPSE/ and \\ http://irsa.ipac.caltech.edu/data/SPITZER/MIPSGAL/} (IRSA).

The search for YSO candidates in Vulpecula relies on the point source catalogs (PSC) generated from the GLIMPSE and MIPSGAL surveys. For GLIMPSE, point sources are extracted using a modified version of the Point-Spread Function (PSF) fitting program DAOPHOT \citep{stetson87}. The GLIMPSE PSC we used is the enhanced data product v2.0 available on the IRSA website. For MIPSGAL, sources are extracted from the 24\mic~maps using a Pixel-Response Function fitting method (Shenoy et al. in preparation) and the APEX software developed at the \emph{Spitzer Science Center}\footnote{http://ssc.spitzer.caltech.edu/postbcd/apex.html}. This catalog is then merged with the GLIMPSE PSC to include IRAC and 2MASS\footnote{The Two Micron All Sky Survey (2MASS) is a joint project of the University of Massachusetts and the Infrared Processing and Analysis Center/California Institute of Technology, funded by the National Aeronautics and Space Administration and the National Science Foundation.} fluxes if they are available. 70\mic~sources are extracted from reprocessed 70\mic~maps (Paladini et al. in preparation) using the \emph{StarFinder} software \citep{diolaiti00}, and matched to the MIPSGAL PSC with a search radius of 9\arcsec.

\subsection{Ancillary data}
\label{subsec:anci}

\begin{figure*} \centering
    \includegraphics[width=0.98\textwidth,angle=0]{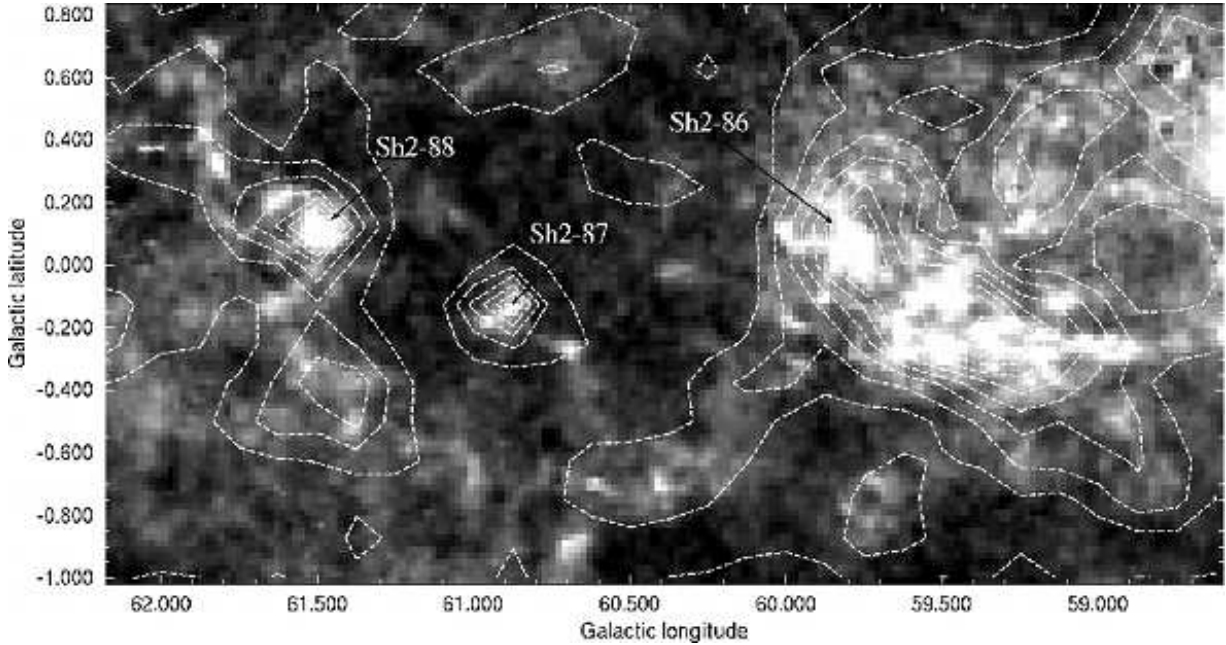}
    \caption{CO contours derived from the velocity-integrated CO map of \citet{dame01}, integrated from 20~to~40$\;$km~s$^{-1}$, overlaid on a visual extinction map of \vul~(Av ranges from 3 to 20$\;$mag on a linear scale). The Sharpless Objects appear as higher extinction regions.\label{fig:Av_CO}}
\end{figure*}

We complement our dataset with visible, sub-millimeter, millimeter and radio data to obtain a broader view of the Vulpecula complex, and to investigate the morphology, star formation history and possible relationship between the \hii regions. We make use of the VLA Galactic Plane Survey (VGPS) data, which consists of \hi line and 21-cm continuum emission data with a resolution of 1\arcmin$\times$1\arcmin$\times1.56$~km~s$^{-1}$ and 1\arcmin~respectively \citep{stil06}. These data are analysed in details in Section~\ref{subsec:h2}. We also use the Virginia Tech Spectral-line Survey (VTSS) which provides arcminute-resolution images of the 6563~\AA~H$\alpha$ recombination line of atomic Hydrogen in the Vulpecula region \citep{dennison98}. S.~Bontemps (private communication, see \cite{schneider06} for details) also provided our team with a 1.2\arcmin-resolution extinction map of \vul~generated from stellar counts and colors derived from the 2MASS point source catalog (see Figure~\ref{fig:Av_CO}). The visual extinction in Vulpecula ranges from 3 to 20 with a distribution peaking at Av$\sim$5.
We also use the CO survey of the Galaxy published in \citet{dame01}. However the relatively low spatial resolution of $\sim$7.5\arcmin~over \vul~prevented a relevant detailed study of the CO emission. We find nonetheless a very good correlation between the CO emission and the visual extinction that both trace the cold and dusty molecular material (cf Figure~\ref{fig:Av_CO}). 
In addition, we used the sub-millimeter data obtained during the second flight of the Balloon-borne Large Aperture Submillimeter Telescope \citep[BLAST,][]{pascale08}, where a 4~square degree region was mapped around Sh2-86 at 250, 350 and 500\mic~\citep{chapin08}. The angular resolution of the BLAST maps are 40\arcsec, 50\arcsec~and 60\arcsec~at 250, 350 and 500\mic, respectively. And finally, we exploit the recently released 1.1\mm~data of the Galactic Plane Survey carried out with Caltech Submillimeter Observatory/Bolocam (Aguirre et al. in preparation) and providing an angular resolution of 31\arcsec.

\section{The Vulpecula OB association}
\label{sec:vulOB1}

\vul~is an active star forming region located in the Galactic plane at a longitude l$\sim$60\degr. In this article we focus on the region centered at $(l,b)=(60.2,-0.2)$ and covering about 6.6~square degrees (see Figures~\ref{fig:Av_CO} and~\ref{fig:vulob1}). \Citet{garmany92} describe \vul~as an oval region about 3.5\arcdeg ~by 1.5\arcdeg~hosting the star cluster NGC$\,$6823. Nearly one hundred hot massive stars are found in the direction of Vulpecula from the catalog of OB~stars compiled by \cite{reed03}. Some of these stars have already been associated with and are responsible for the ionization of three bright \hii regions in \vul, namely Sh2-86, 87 and 88 \citep{sharpless59}. Figure~\ref{fig:vulob1} presents the mid-infrared morphology of \vul~and the location of the OB stars. 

The VGPS 21-cm continuum emission covering \vul~reveals two bright circular compact sources ($\sim$1.3\arcmin~in diameter) coincident with Sh2-87 and Sh2-88, and an extended dimmer region (27\arcmin$\times$12\arcmin) reminiscent of the infrared morphology of Sh2-86. The supernova remnant SNR~G59.5+0.1 \citep{taylor92} is also clearly visible in the radio continuum data at $(l,b)=(59.59,0.1)$, its measured diameter is 15\arcmin~(cf Section~\ref{subsec:snr}). 
Our analysis of the VGPS \hi line data indicates that the three \hii regions are neighbors (cf Section~\ref{subsec:h2}), as was previously noted by \cite{turner86}, \cite{ehlerova01} and \cite{cappa02}. Distance determinations for the three Sharpless objects and their exciting stars range from 1.5~to~3.2~kpc \citep{fich84,barsony89,guetter92,brand93,massey95,deharveng00,hoyle03,kharchenko05,bica08}. Following the arguments from \citet{chapin08}, we adopt a common distance of 2.3~kpc for the three \hii regions.

In  the catalog of \citet{reed03} we find about 30~OB stars in the direction of Sh2-86 and Sh2-87, in what we identify as the \emph{pillars region} (these stars might be responsible for the existence of pillar-structures, see Section~\ref{subsubsec:pillars}). We check their possible association with \vul~by computing their distance moduli based on the $M_V$-Spectral type relation and the intrinsic colors given by \citet{LB82}. We assume a ratio of total-to-selective absorption $R_{\rm V}=3.1$ typical of the diffuse ISM, and we associate a visual extinction to each star based on our extinction map of \vul. Out of the 16~stars for which spectral type and magnitude information are available in the \citeauthor{reed03} catalog, 12~fall between 1.3~and 2.3~kpc with a mean distance from the Sun of $1.91 \pm 0.4$~kpc. We find similar distances using intrinsic colors and absolute magnitudes given by \citet{johnson58} and \citet{wegner06} respectively. Although the uncertainties are significant, which is mostly due to the large uncertainty on the values of the absolute magnitudes, our distance estimate for these stars is consistent with an association with the Vulpecula complex.
%

\subsection{\hii regions}
\label{subsec:h2}

Sh2-86 is an extended \hii region about 40\arcmin~wide \citep{sharpless59} located in the southern part of Vulpecula (Figures~\ref{fig:Av_CO},~\ref{fig:vulob1} and~\ref{fig:sh86_i3i4m1} in the appendices). It is excited by the open cluster NGC~6823, which is part of the OB association \vul~\citep{massey95,bica08}. \citet{pigulski00} estimate the age of the star cluster at $3\pm1$~Myr old. In the atomic gas distribution derived from the VGPS data, Sh2-86 appears as an oblong hole at velocities between 26 and 31~km~s$^{-1}$. We find that the dimension of the shell is approximately 60\arcmin~along the Galactic plane and 25\arcmin~across it. In the optical, the south-eastern part of Sh2-86 is very contrasted. It exhibits silhouetted pillar-like structures pointing towards NGC~6823, emission and reflection nebulae as well as filamentary structures. The infrared emission is also contrasted in this region tracing the complex interplay between the UV radiation and the surrounding dusty interstellar medium. \\
\citet{chapin08} also report the detection of 49~compact sub-millimeter sources associated with Sh2-86 using the Balloon-borne Large Aperture Submillimeter Telescope (BLAST). The presence of these clumps, with masses ranging from 14 to 700~M$_{\sun}$, indicates the capability of Sh2-86 to form massive stars. 

Sh2-87 and Sh2-88 are also \hii regions initially discovered by \citet{sharpless59}. These regions are active sites of star formation as indicated by the presence of H$_2$O maser line emissions and bipolar outflows \citep{barsony89, deharveng00}. Bolocam data show bright sources at the location of Sh2-87 and Sh2-88 indicating the presence of cold core candidates. Our analysis of the VGPS data reveals that the morphology of the \hi shell around Sh2-87 at velocities between  22 and 26~km~s$^{-1}$ is very similar to that of the 24\mic~emission. The \hii region appears as a hole in the \hi distribution as the hydrogen is ionized by an embedded B0 star \citep{felli81}. A compact source seen in absorption sits at the center of Sh2-87, its position is coincident with the location of the peak emission at 24\mic~$(l,b)=(60.88, -0.13)$. \citet{barsony89} and \citet{xue08} give a detailed description of this \hii region. 

Sh2-88 is a diffuse nebula of diameter $\sim$20\arcmin~located at $(l,b)=(61.45, 0.34)$, which is excited by the O8 star BD~+25\degr3952 \citep{cappa02}. The star formation activity taking place in Sh2-88 occurs in a couple of nebular knots identified by \citet{lortet74}, Sh2-88A and Sh2-88B, located 15\arcmin~southeast of Sh2-88. These knots are responsible for the bright 24\mic~emission visible in Figure~\ref{fig:vulob1} at $(l,b)=(61.47, 0.1)$. Sh2-88B is actually the brightest of the three Sharpless objects at mid-IR wavelengths. Figure~\ref{fig:sh88b_i1i3i4} in the appendices shows a composite image of Sh-88B in IRAC bands. It consists of a compact cometary \hii region and an ultracompact (UC) \hii region. \citet{deharveng00} present a detailed analysis of Sh2-88B and its stellar content. At radio wavelength, the association of Sh2-88 with a \hi shell is somewhat more difficult than in the other two cases as the edges of the shell are not as well defined. We detect however a faint hole at velocities between 22 and 26$\,$km~s$^{-1}$ as well as  a compact source seen in absorption in the center of Sh2-88B. 

\begin{figure*}\centering
    \begin{tabular}{r}
    \includegraphics[width=0.873\textwidth,angle=0]{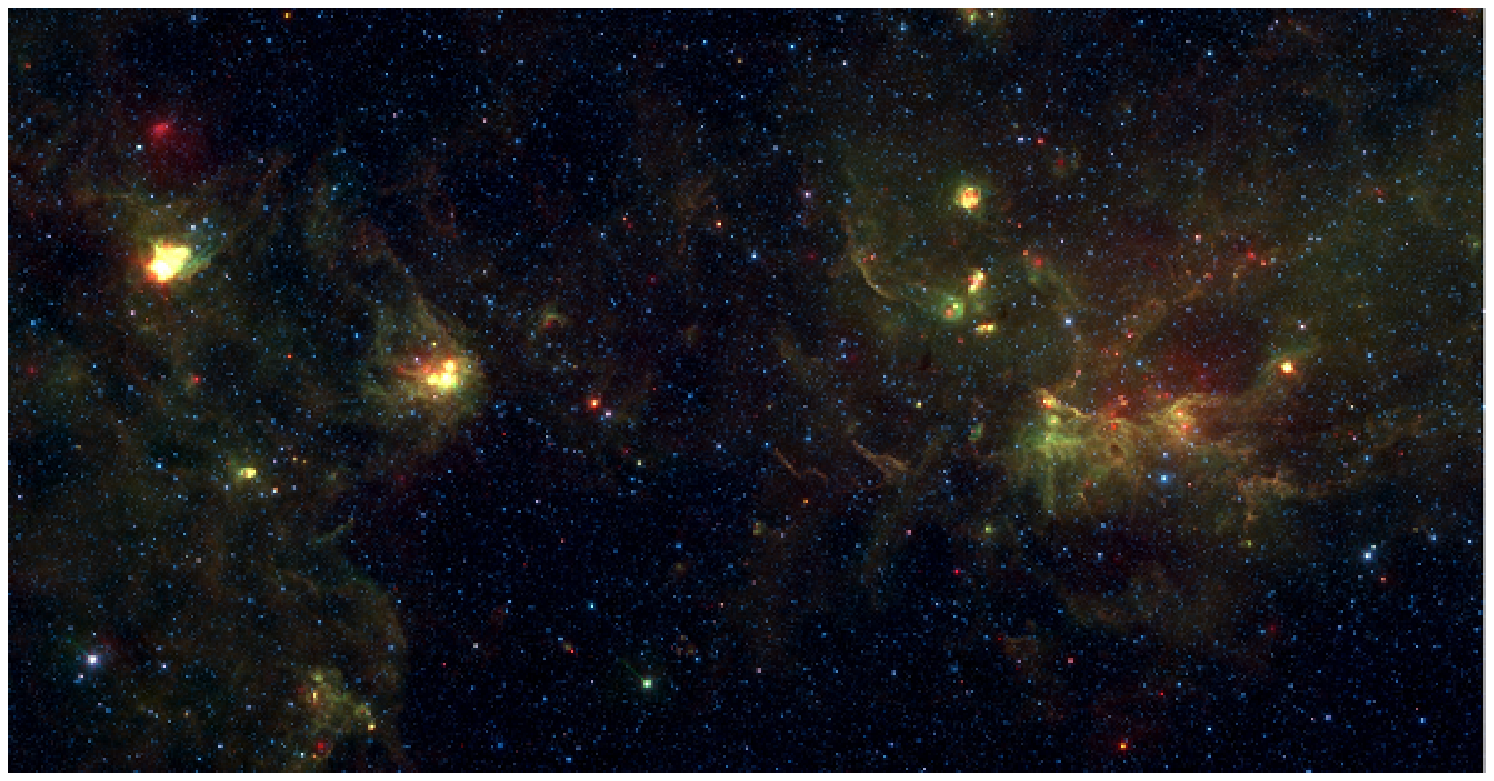}\\
    \includegraphics[width=0.95\textwidth,angle=0]{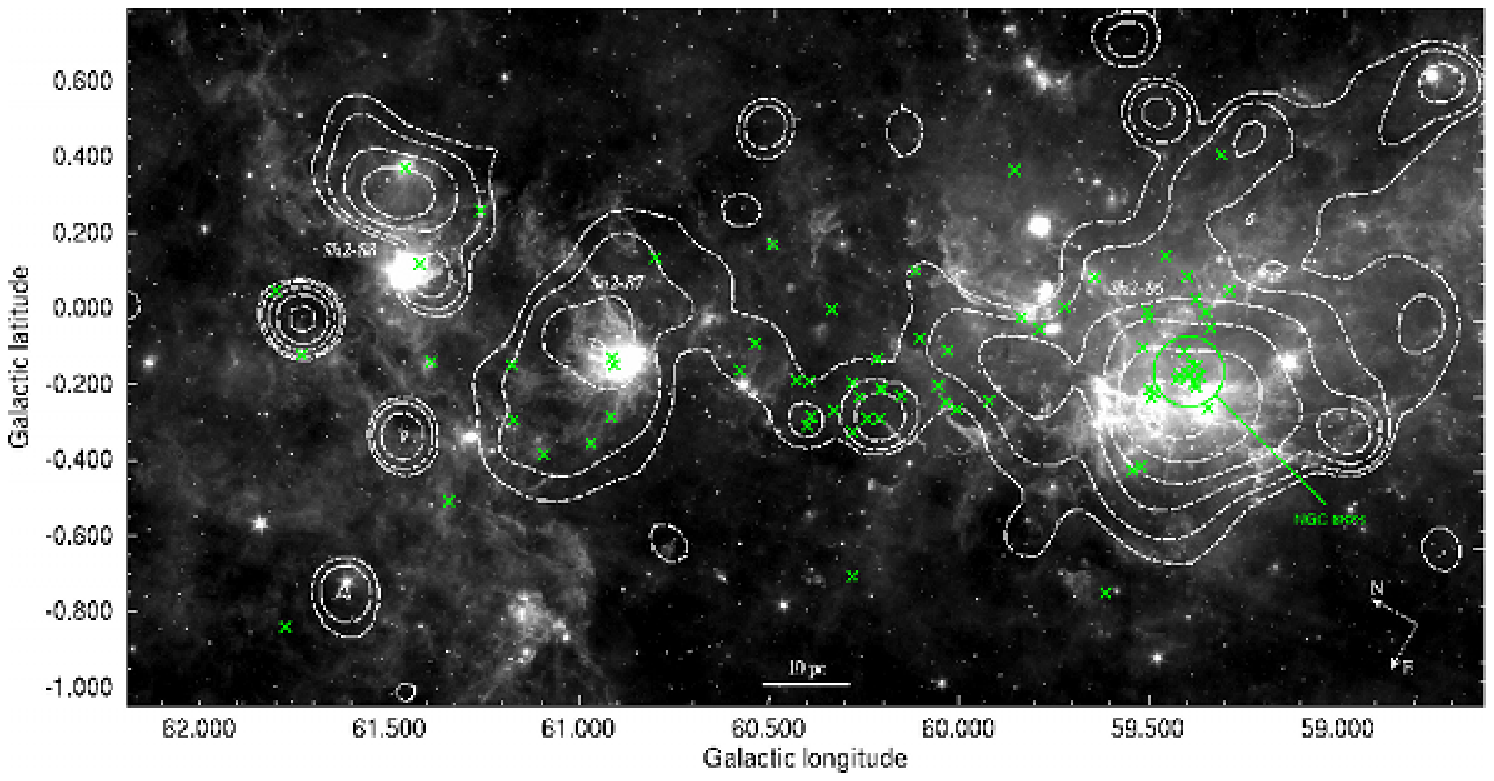}
    \end{tabular}
    \caption{The top panel shows a composite image of \vul~(3.6, 8.0 and 24$\;\mu$m are color-coded as blue, green and red respectively). This image was assembled by R. Hurt and T. Robitaille based on GLIMPSE and MIPSGAL data. The bottom panel represents H$\alpha$ contours from the VTSS overlaid on the MIPSGAL~24$\;\mu$m image. The H$\alpha$ emission associated with Sh2-88 peaks at $(l,b)=(61.45, 0.30)$ while the bright mid-IR emission located 15\arcmin~south-east of the diffuse nebula is the knot identified in the text as Sh2-88B. Green crosses indicate the location of the 88~OB$\;$stars present in this 3.7\degr$\times$1.8\degr~region.\label{fig:vulob1}}
\end{figure*}

\subsection{Pillar structures}
\label{subsubsec:pillars}

We have discovered several pillar-like structures on either side of the Galactic equator between Sh2-86 and Sh2-87 ($59^\circ\!.5 < l < 61^\circ$). These objects are similar to the archetypal \emph{pillars of creation} found in M16 \citep{hester96,urquhart03}. Such pillars are usually associated with recent episodes of star formation as the winds and radiation emanating from young massive stars are responsible for sculpting these elongated \emph{elephant trunks} out of the surrounding molecular material. An accepted mechanism for the formation of the pillars is the slow photoevaporation of a pre-existing molecular clump shadowing a tail of more diffuse gas, but other mechanisms have been proposed based on hydrodynamical instabilities for instance. \citet{spitzer54}, \citet{bertoldi89}, \citet{lefloch94} and \citet{carlqvist03} present analytical models for the formation of pillar-like structures, and \citet{miao06}, \citet{mizuta06} and \citet{gritschneder09} carry out numerical simulations of the formation and evolution of such objects. According to these studies, the pressure at the surface of the pillars due to the strong external radiation appears to trigger the formation of new stars, which is confirmed by recent observations \citep[e.g.][]{sugitani02,reach04,bowler09,reach09}

\begin{figure*}\centering
    \includegraphics[width=0.98\textwidth,angle=0]{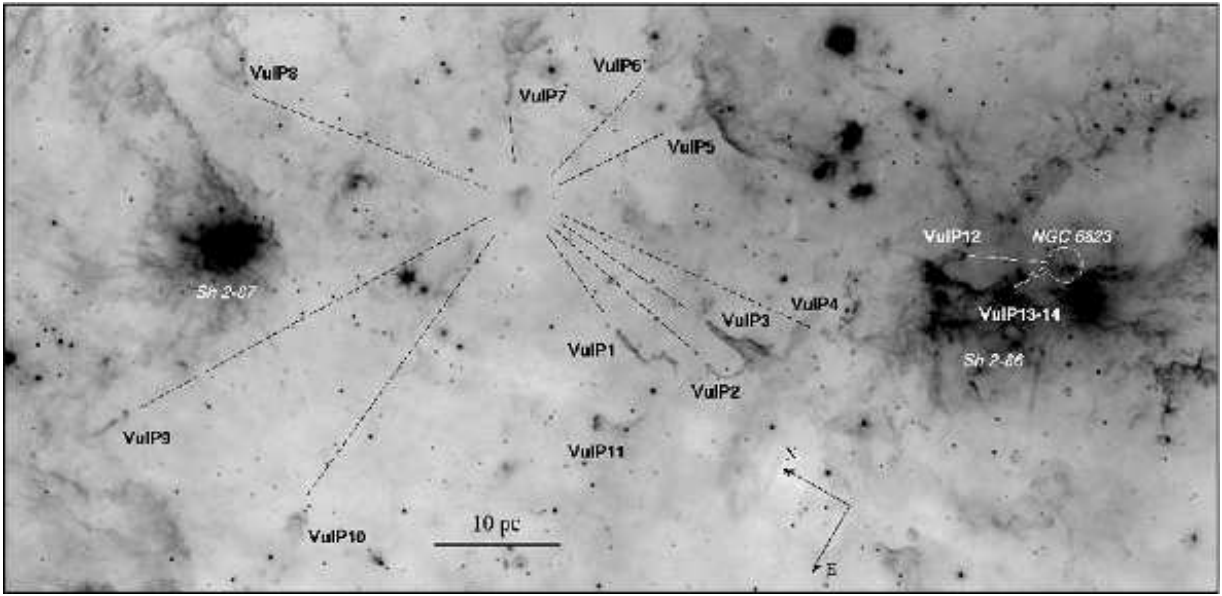}
    \caption{Identification of pillar structures in \vul~on a MIPSGAL$\,$24\mic~image. The grey scale is stretched to increase the contrast of the pillars. \emph{VulP1} to \emph{VulP10} seem to point toward the same direction. Extended concentric structures are noticeable around the central position up to a radius of $\sim10\,$pc. \emph{VulP12} to \emph{VulP14} are associated with the young star cluster NGC~6823. \label{fig:pillar_24mu_ID}}
\end{figure*}

Figure~\ref{fig:pillar_24mu_ID} presents the 24\mic~image of the pillars found in Vulpecula, and gives the nomenclature to identify individual objects. The three pillars \emph{VulP12-13-14} were already known prior to these observations \citep{chapin08}; they are located northeast of NCG~6823 and are obviously associated with NGC~6823 as they point towards the star cluster. However the 11 pillars \emph{VulP1} to \emph{VulP11} have never been reported in the literature before this study. Their association with \vul~is not as straightforward as in the case of NGC~6823 since no obvious source could be identified as their sculptor. 
The OB stars mentioned previously are certainly good candidates, for instance the star HD~186746 (spectral type B8Ia) is located right above the tip of the pillar \emph{VulP1} in projection (Figures~\ref{fig:vulob1} and~\ref{fig:pillar_24mu_ID}). 
Nevertheless, 10~of the discovered pillars, out of~11, seem to point towards the same faint diffuse nebulosity located at $(l,b)=(60.37, -0.04)$, which is not coincident with any known OB stars. We argue that the chance for ten randomly distributed pillars to point towards the same object is negligible\footnote{Assuming we measure the direction of the pillars with an accuracy of $\sim2$\degr, the probability is lower than $10^{-11}$.}, thus the discovered pillars \emph{VulP1}-\emph{VulP10} are most certainly associated with each other. The central nebulosity might host the object responsible for the formation of the pillars. In addition, a diffuse faint circularly symmetric structure centered on the same nebulosity is discernible in the background diffuse emission. Bright objects at 24\mic~also seem to delineate this faint structure at a radius of $\sim15$\arcmin~centered at $(l,b)=(60.37, -0.04)$. These morphological features suggest that a single event in the past may have molded the pillars region. 

We analyze the radio data in the pillars region to look for a possible association between these structures and the OB~association. Out of the 15~pillars identified in Figure~\ref{fig:pillar_24mu_ID}, only 3 are detected in the VGPS \hi line data. The pillar \emph{VulP10} presents a deep absorption feature at 27-33~km~s$^{-1}$ as well as smaller features centered at 8, 13 and 20~km~s$^{-1}$. These \hi features are marginally resolved into two close compact sources located at the base of the pillar. Since these features are quite different from the IR morphology of \emph{VulP10}, we cannot exclude a fortuitous association with a coincident radio source seen in projection. 

\emph{VulP4} and \emph{VulP5} appear to have similar characteristics, they present comparable morphologies in the \hi line data and in the \emph{Spitzer} bands. They both emerge from the molecular cloud to be exposed to the ionizing radiation. These pillars are located at the edge of \hi shells, and their silhouette is clearly identified at velocities 21-23~km~s$^{-1}$, which places them at roughly the same distance as the three Sharpless objects for which $V_{LSR}=22-31$~km~s$^{-1}$. 

VGPS continuum emission shows a bright two-lobed compact source located at the base of the pillar \emph{VulP10}, consistent with the two blobs observed in the \hi line data. No other pillars are detected in the 21-cm continuum.

\begin{figure*}\centering
    \begin{tabular}{c}
    \includegraphics[width=0.85\textwidth,angle=0]{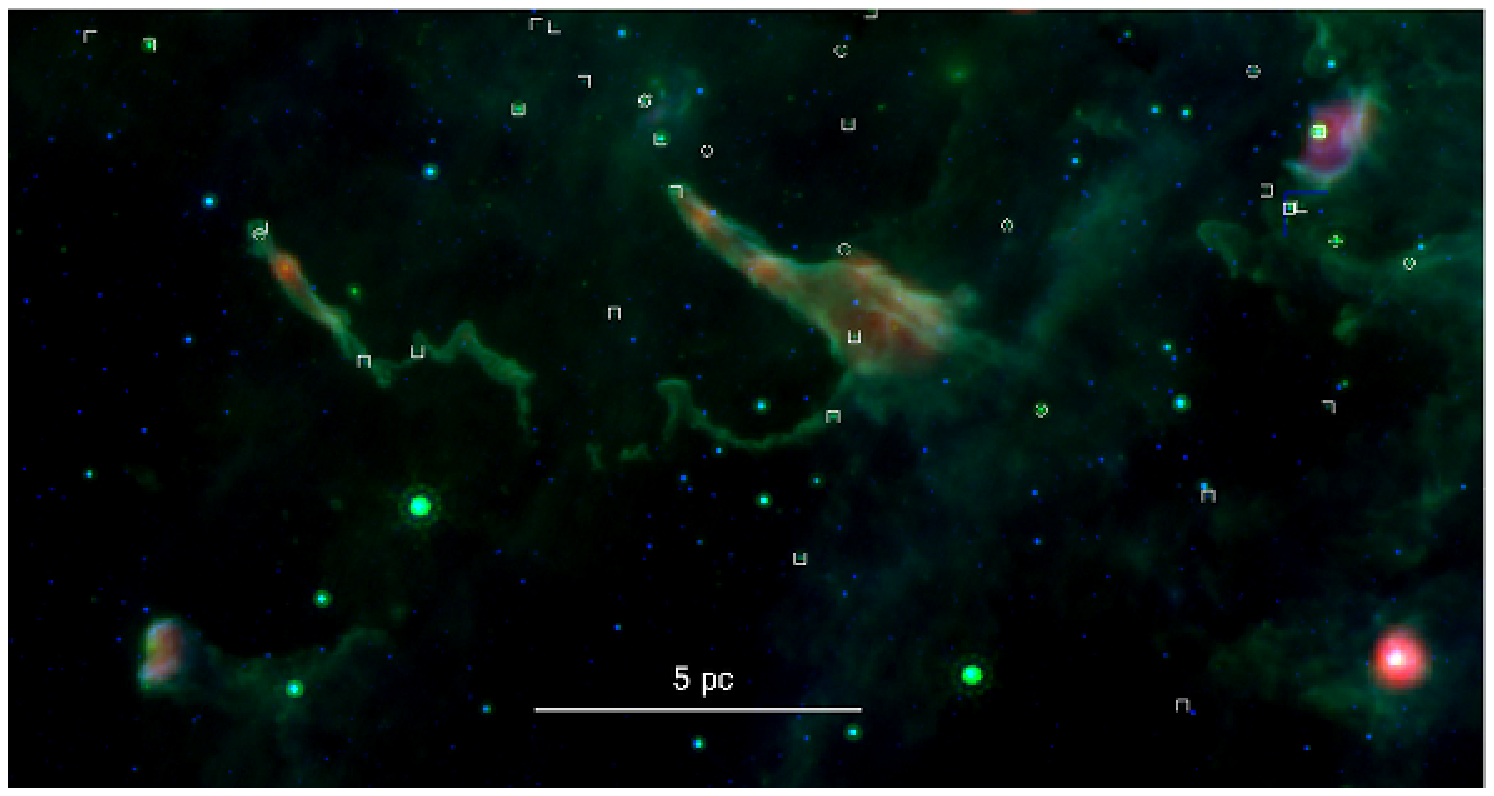}\\
    \includegraphics[width=0.85\textwidth,angle=0]{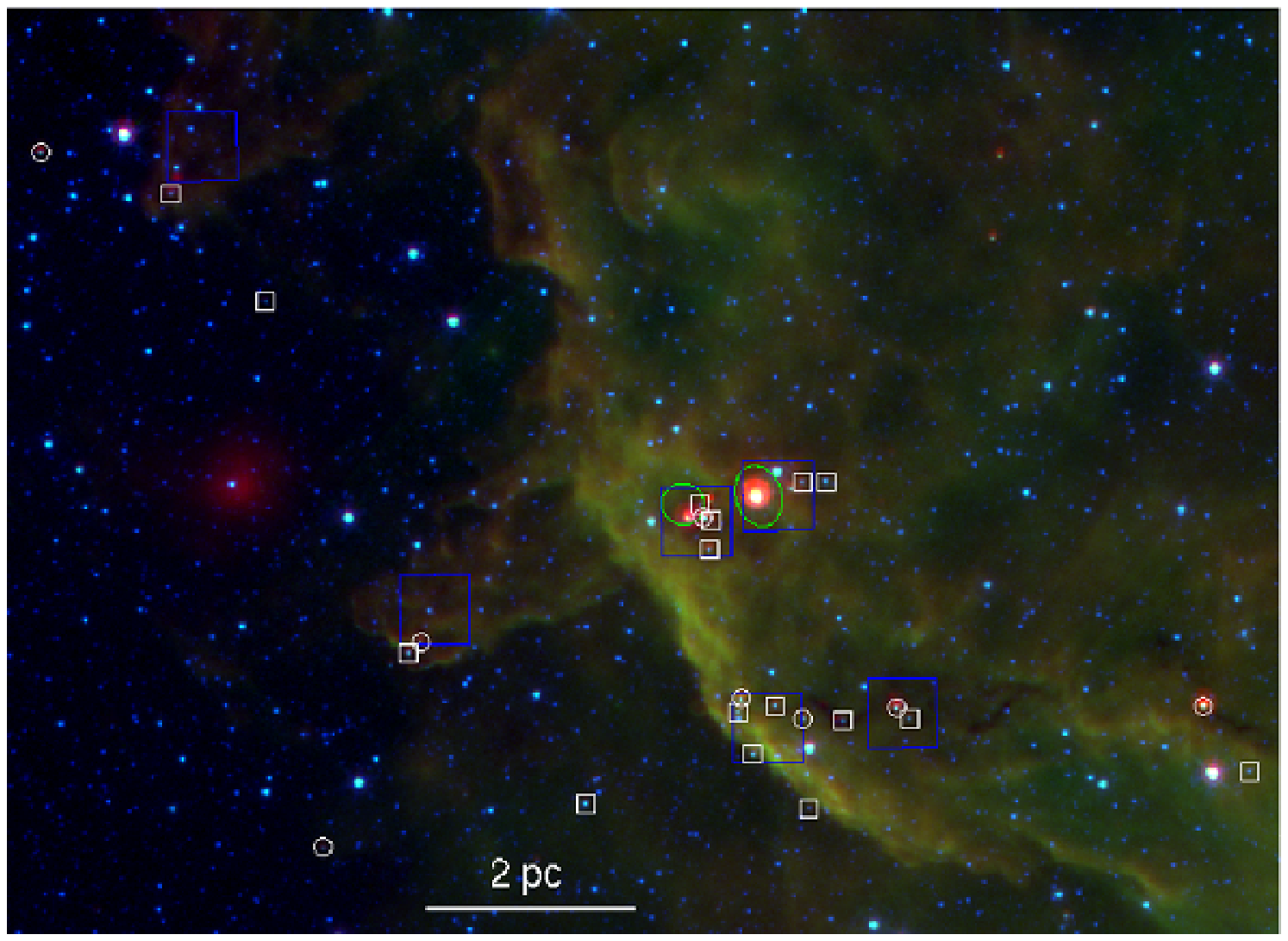}
    \end{tabular}
     \caption{\emph{Spitzer} 3-color images of some pillars discovered in \vul. YSO candidates are marked as white small squares and circles, large blue boxes show BLAST sources and green ellipses indicate Bolocam sources. \emph{Top panel}: Five pillars identified as \emph{VulP11} and \emph{VulP1} to \emph{VulP4} in Figure~\ref{fig:pillar_24mu_ID} (from left to right). Blue, green and red are 8, 24 and 70\mic~images, respectively. \emph{VulP1} and \emph{VulP2} exhibit bright 70\mic~sources in their core. \emph{Bottom panel}: \emph{VulP5} and  \emph{VulP6}. Blue, green and red are respectively 4.5, 8.0 and 24\mic~images. InfraRed Dark Clouds (IRDCs) are visible in the pedestal of the pillar \emph{VulP5}. YSO candidates are aligned along IRDCs.  \label{fig:pillars_color}}
\end{figure*}

Tables~\ref{tab:pillars_coord} and~\ref{tab:pillars_detail} in the appendices summarize the geometrical and morphological information extracted from the \emph{Spitzer} observations for individual pillars. Of particular interest are the bright ($\sim2$~Jy) compact sources found at 70\mic~in the core of the pillars \emph{VulP1} and \emph{VulP3} (see Figure~\ref{fig:pillars_color}). These embedded sources are faint at 24\mic~($\sim1$~mJy) and have no counterparts in IRAC and 2MASS bands. Even though star formation is expected to occur preferentially at the tip of such structures, these red sources are likely sites of massive star formation. Longer wavelength\footnote{The BLAST data do not cover these pillars, and the 1.1~mm Bolocam data, which is available for these pillars, is not as good a diagnostic as the sub-millimeter data.} observations would be necessary to identify the nature of these sources. Figure~\ref{fig:pillars_color} shows the pillars  \emph{VulP5} and \emph{VulP6} and their associated red sources. YSO candidates are located at the tip of each pillar, and also in the pedestal of \emph{VulP5} along some Infrared Dark Clouds (IRDC). Sub-millimeter BLAST sources are also located at the tip of the pillars and along the same IRDCs. Two Bolocam sources are also located along one of the IRDCs.

\section{The IR-excess emission of Young Stellar Objects}
\label{sec:YSO}

Stars form from the gravitational collapse and fragmentation of giant molecular clouds in the ISM. The contraction of the cold gas leads to the formation of a dense rotating core radiating its thermal energy in the millimeter and far-IR regime. As the bulk of the initial cloud mass is falling towards the center of the core, a protostar emerges with a gaseous and dusty accretion disk rotating around it. The peak emission of such a young stellar object shifts towards shorter wavelengths revealing a bright source in the mid-IR. The circumstellar disk is then dissipated via accretion, planet formation or evaporation. Further reading on star formation mechanisms can be found in e.g. \cite{terebey84, adams87, andre93, andre00,mckee07}.

Recent works have shown that the wavelength coverage of \emph{Spitzer} instruments is well suited for observing the mid-IR excess emission emanating from the warm circumstellar material around Young Stellar Objects \citep[e.g.][]{allen04,harvey07,koenig08,guieu09}. Based on the mid-IR extinction law \citep{lutz99,indebetouw05,flaherty07}, \citet{gutermuth08} furthermore argue that the most reliable IRAC-based criterion for identifying YSO candidates is the [4.5]-[5.8] color. Indeed the flattening of the extinction curve observed between the IRAC bands 2 and 4 (4.5 to 8\mic) reduces the degeneracy between selective interstellar extinction and intrinsic IR excess. 

Our approach to identify YSOs in \vul~is to exploit the method developed by \cite{gutermuth08}  based on \emph{Spitzer} colors and magnitude cuts to identify sources with IR-excess emission. We apply this method to the Point Source Catalogs (PSC) compiled by the GLIMPSE and MIPSGAL teams (Section~\ref{subsec:obs_cat}) to take full advantage of the good photometric quality and the high reliability of these data. The direct consequence of this approach is that no extended sources will be considered in our study even though they exhibit significant IR-excess emission. To mitigate this caveat, we compare the angular resolution of \emph{Spitzer} with the typical angular size of circumstellar envelops and disks seen at 2.3~kpc from the Sun. Assuming the telescope optic is diffraction limited\footnote{From the \emph{Spitzer} Observer's Manual available at http://ssc.spitzer.caltech.edu.} down to 5.5\mic, the resolution element at 3.6 and 24\mic~corresponds to 3500 and 16000~AU at 2.3~kpc respectively. The spatial extend of the Young Stellar Objects we are looking for does not exceed these limits. For instance, \citet{vicente05} measured the size distribution of circumstellar disks in the Trapezium cluster and found disk sizes ranging from 100 to 400~AU, an order of magnitude smaller than the resolution element at 3.6\mic. 
Although protostellar envelops have a larger extend, i.e. a few thousand~AU according to \citet{bonnell96} and \citet{jorgensen02}, and assuming they can only be detected at long wavelengths where the resolution element is larger, we expect the number of resolved YSOs in \vul~to be relatively small. 
However a fraction of the YSOs in their early phases of evolution might be resolved by \emph{Spitzer}, even at 24\mic, and those large protostars would be excluded from our analysis. 

\subsection{Initial catalog}
\label{subsec:initial_cat}

We first compile a catalog of IRAC point sources based on the highly reliable Point Source Catalog available from the IRSA website as a GLIMPSE$\,$I v2.0 enhanced data product\footnote{http://www.astro.wisc.edu/glimpse/glimpse1\_dataprod\_v2.0.pdf}. All sources found in this catalog with galactic coordinates in the range $58.6\degmath<l<62.2\degmath$ and $-1.0\degmath<b<0.8\degmath$ are considered ($\sim3.28\times 10^5$~sources). Fluxes are converted to magnitudes using the zeropoints given in Table~\ref{tab:zeropoints}. To ensure good photometry and avoid \emph{flux stealing} between adjacent sources as noted by \cite{robitaille08}, we require that valid point sources have magnitude errors lower than~0.2$\,$mag, and that their close source flag\footnote{Following the convention of the GLIMPSE$\,$I v2.0 enhanced data products, csf=0 implies that no sources in the Archive Catalog are within 3\arcmin~of the source.} (csf) is zero. We also apply specific cuts on IRAC magnitudes to achieve high detection reliability, the magnitude limits we adopt are 14.2, 14.1, 11.9 and 10.8~mag at 3.6, 4.5, 5.8 and 8.0 $\mu$m respectively. According to the GLIMPSE$\,$I Assurance Quality Document\footnote{http://www.astro.wisc.edu/glimpse/GQA-master.pdf}, these limits ensure a detection reliability above 98\% in all four IRAC bands. The reliability of the faintest sources however might vary across the \vul~region as it depends on the background brightness structure and the local source density.

This intermediate catalog is then merged with the MIPSGAL~24\mic~Point Source Catalog (Shenoy et al. in preparation) with the constraint that the maximum separation between two matching sources is 3\arcsec. 
Once again to ensure a good photometry, we require that the magnitude of 24\mic~sources and their associated errors are constrained. These constraints however are applied at a later stage of the selection procedure (see Section~\ref{subsec:YSO_selection}).

Out of nearly $3.3\times10^5$ sources present in the GLIMPSE catalog within our search area, only $\sim1.47\times10^5$ bypassed the initial magnitude cuts. Most of the exclusions were due to noisy detections in bands~3 and~4 of IRAC. 

In the end, the initial catalog from which we search for YSO candidates contains point sources with well measured photometry and with at least one detection in IRAC bands, plus the associated 2MASS and MIPSGAL~24\mic~fluxes if they exist.

\begin{deluxetable}{l@{\extracolsep{25pt}}c@{\extracolsep{10pt}}c@{\extracolsep{10pt}}c@{\extracolsep{25pt}}c@{\extracolsep{10pt}}c@{\extracolsep{10pt}}c@{\extracolsep{10pt}}c@{\extracolsep{25pt}}c@{\extracolsep{10pt}}c} \centering
	\tablecolumns{10}
	\tabletypesize{\small}
	\tablewidth{0pt}
	\tablecaption{Zeropoints used for Flux-to-Magnitude Conversions.\label{tab:zeropoints}}
	\tablehead{\colhead{}& \multicolumn{3}{c}{2MASS} & \multicolumn{4}{c}{IRAC} & \multicolumn{2}{c}{MIPS} \\
	 \cmidrule(r){2-4}  \cmidrule(r){5-8}  \cmidrule(r){9-10}
 	Band & J & H & K & [3.6] & [4.5] & [5.8] & [8.0] & [24] &[70]}
	\startdata
	Zeropoint [Jy] & 1594 & 1024 & 666.7 & 280.9 & 179.7 & 115.0 & 64.13 & 7.17&0.778
	\enddata
\tablecomments{2MASS zeropoints are from \cite{cohen03}, IRAC from \cite{reach05} and MIPS from \cite{rieke08}.}
\end{deluxetable}

\subsection{Mid-IR selection of YSO candidates}
\label{subsec:YSO_selection}

We start our search for YSO candidates in the initial catalog by removing extragalactic \emph{contaminants} that might be misidentified as YSOs. Using the selection criteria presented in the appendices of \citet{gutermuth08}, which are based on a complementary analysis of the Bootes field IRAC data \citep{stern05}, we identify and reject 6~sources dominated by Polycyclic Aromatic Hydrocarbons (PAH) feature emission, likely star-forming galaxies or weak-line active galactic nuclei (AGN). No sources matching the colors and magnitudes criteria for broad-line AGNs were found. Considering a search area of 6.5~square degrees, we find a density of extragalactic sources in Vulpecula slightly lower than one per square degree. This is comparable to the density of $\sim$0.5~galaxy per square degrees found in the zone of avoidance by \citet{marleau08} using GLIMPSE and MIPSGAL data. We further filter the catalog according to \citeauthor{gutermuth08} criteria and reject 8~sources that have a large 4.5\mic~excess emission consistent with molecular hydrogen line emission found in regions where high-velocity outflows interact with the cold molecular cloud. The remaining sources are presumably of stellar origin.

All sources with the following color constraints are considered likely YSOs: 
$$[4.5]-[8.0] > 0.5,$$
$$[3.6]-[5.8] > 0.35,$$
$$[3.6]-[5.8] \le 3.5\times([4.5]-[8.0])-1.25,$$
Out of the $\sim2.38\times10^4$ sources possessing all four IRAC magnitudes, we identified 820~YSO candidates based on their IR-excess emission. We further classify the selected objects according to their infrared spectral index $\alpha_{IR}=\partial\, log(\lambda F_\lambda)/\partial \,log(\lambda)$ as defined by \citet{lada87}. We specifically compute the spectral index $\alpha_{IRAC}$ as the slope of the Spectral Energy Distribution (SED) measured from 3.6 to 8.0\mic. Objects with $\alpha_{IRAC} > -0.3$ have a flat-ish or rising SED indicating the presence of a cold dusty envelop infalling onto a central protostar, these are designated class$\;$0/I YSOs. Objects with $-0.3 > \alpha_{IRAC} > -1.6$ are classified as class$\;$II YSOs, these are pre-main-sequence stars with warm optically thick dusty disks orbiting around them. We define class$\;$III objects as having  $-1.6> \alpha_{IRAC} > -2.56$: these stars have cleared most of their circumstellar environment, their near-IR emission is mostly photospheric but they may present some excess emission above 20\mic. Such objects are dubbed \emph{anemic} disks by \citet{lada06}. Finally, objects with $\alpha_{IRAC} < -2.56$ are stars with photospheric emission only, this spectral index corresponds to the slope of the stellar photosphere SED in the Rayleigh-Jeans domain. 

\begin{figure*}\centering
    \begin{tabular}{ll}
    \includegraphics[width=0.48\textwidth,angle=0]{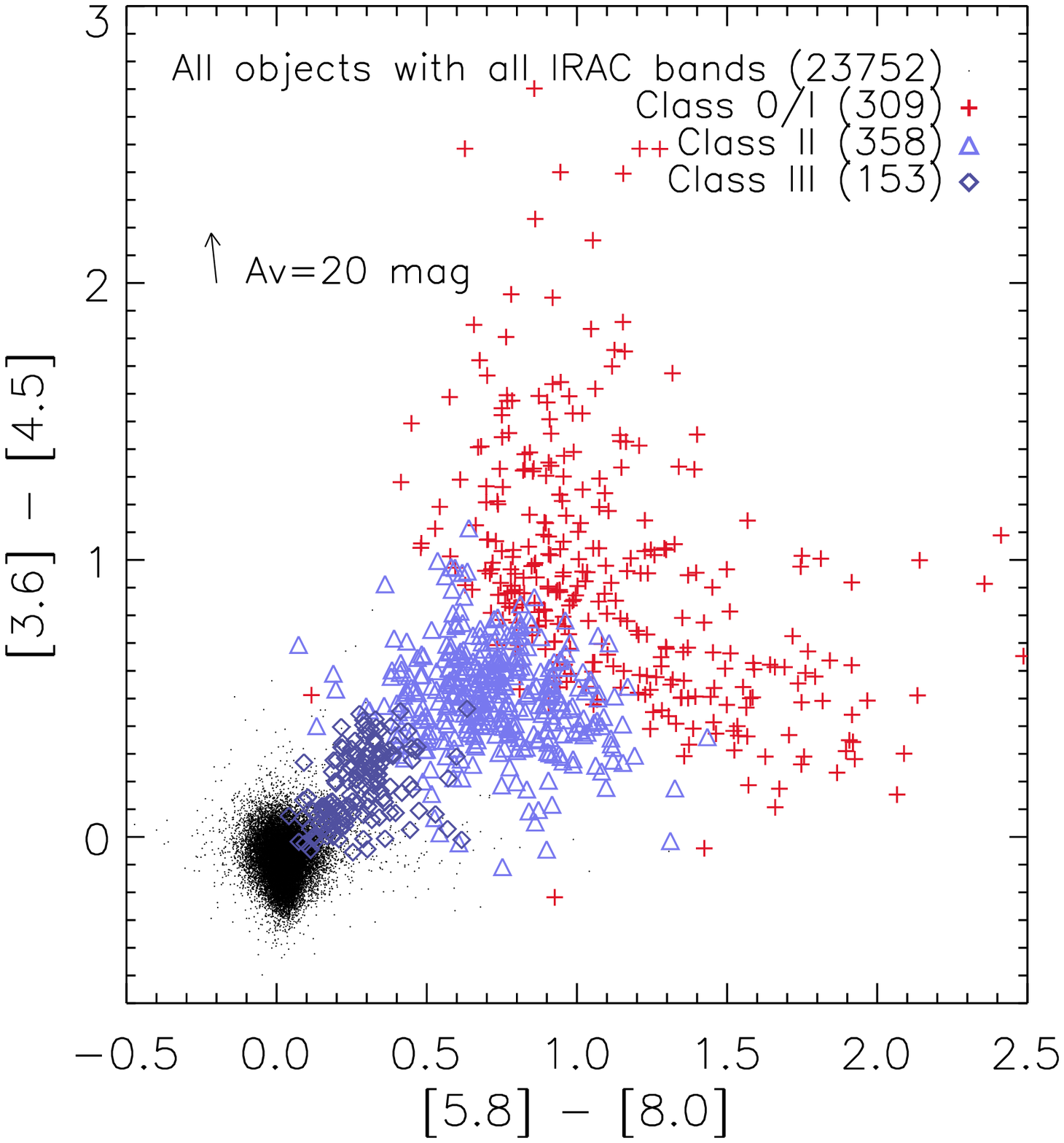} &
    \includegraphics[width=0.48\textwidth,angle=0]{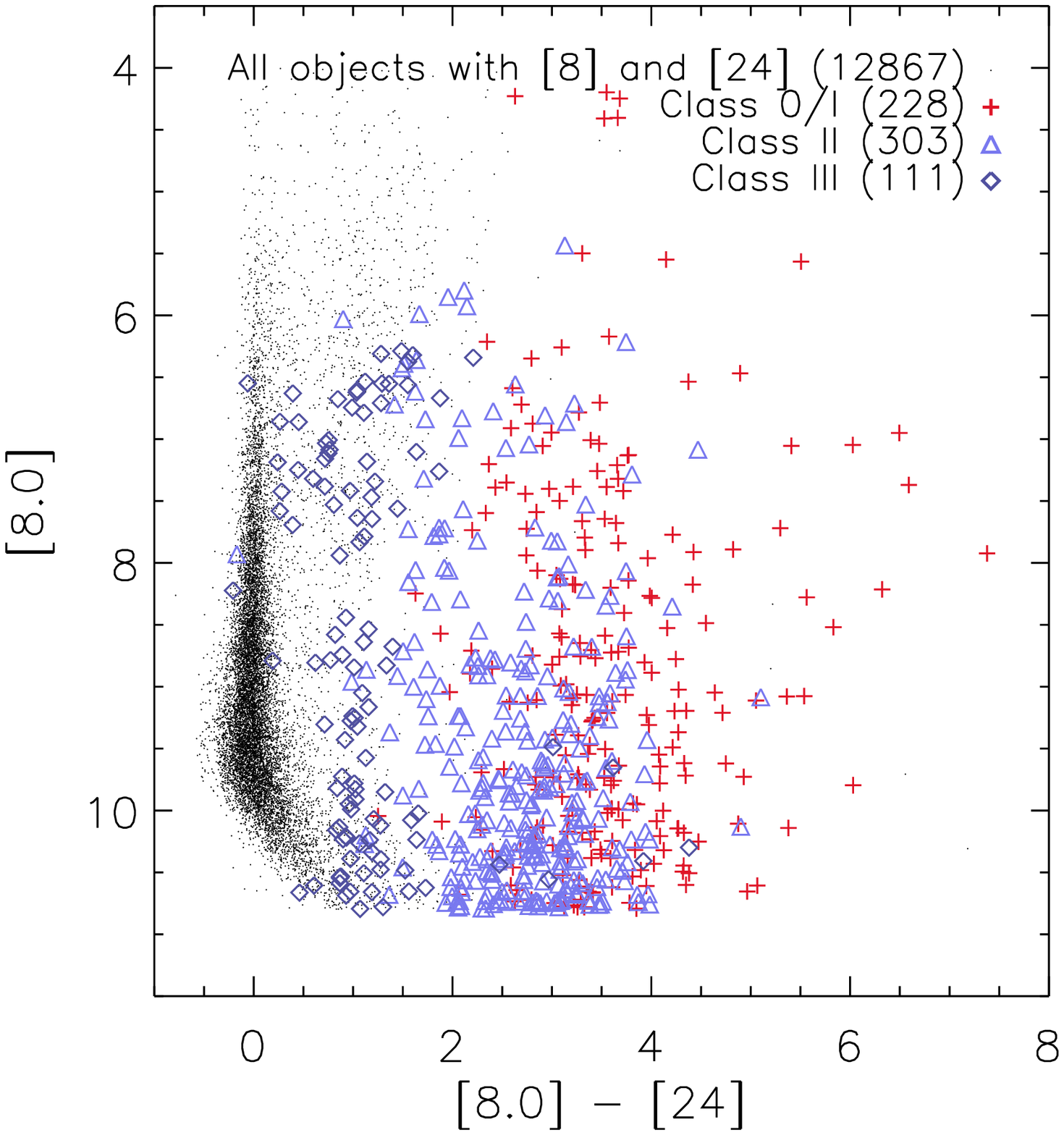}
    \end{tabular}
    \caption{\emph{Left panel}: IRAC color-color diagram of point sources found in the Vulpecula region. Stars with photospheric emission are marked as black dots in the figure, they have null colors in IRAC bands so they aggregate around point (0,0). Color- and symbol-coded YSO candidates protrude from the bulk of the stellar sources towards the positive color quadrant due to their IR-excess emission. The reddening vector is derived from the extinction law of \citet{flaherty07}. \emph{Right panel}: Color-Magnitude diagram of point sources based on MIPS and IRAC bands.  (A color version of this figure is available in the online journal.) \label{fig:ccdiag_irac}}
\end{figure*}

Note that this YSO classification is based on the classification scheme proposed by \citet{greene94}, except that the Flat Class ($0.3>\alpha>-0.3$) is included into the class$\;$0/I which represents the population of protostars with infalling envelopes. \citet{calvet94} indeed showed that Flat Class YSOs could be interpreted as infalling envelopes. Figure~\ref{fig:ccdiag_irac} presents a color-color diagram based on IRAC data only, the different symbols indicate the classes associated with each YSO candidate identified in our analysis.\\

The next step is to exploit the extra information contained at longer wavelengths to confirm, or reclassify, the YSO candidates identified solely from their IRAC colors. We first check that the SEDs of all YSO candidates continue to rise from~8 to~24\mic. We also look for objects previously classified as photospheric sources that exhibit bright 24\mic~fluxes ($[5.8]-[24]>1.5$). These sources are thought to be \emph{transition disks}, i.e. class$\;$II YSOs with significant dust clearing from their inner disks \citep{calvet02,dalessio05}, these objects are of particular importance for understanding the evolution of circumstellar disks around young stars. We also check for protostar misclassifications due to extreme visual extinction levels. \citet{gutermuth08} suggests that if a protostar that has MIPS detection does not meet the criterion $[5.8]-[24]>4$ (if they possess $[5.8]$ photometry) or $[4.5]-[24]>4$,  then it is likely a highly reddened class$\;$II YSO. Finally, we require that any sources lacking detections in some IRAC bands yet having bright 24\mic~fluxes ($[24]<7$ and $[X]-[24]>4.5$~mag, where $[X]$ is the photometry for any IRAC detection available in our catalog) have to be added to the list of likely YSOs, and to be classified as highly embedded protostars.
We apply the 24\mic-based color constraints on YSO candidates only if their MIPS detection reliability is $>95$~\% (S.~Shenoy, private communication); thus we require that, for the MIPS photometry to be relevant for the YSO selection, the 24\mic~magnitude is $[24]<8.6$~mag and that its associated error is $\sigma_{[24]} <0.2$~mag.\\

We tested the ability of the method at finding genuine YSOs on a well studied star forming region, RCW79 \citep{zavagno06}, and it proved to be adequately reliable and efficient (Zavagno, private communication). Furthermore, this method has been successfully applied to study embedded stellar clusters in NGC~1333 \citep{gutermuth08}, the star formation activity in the \hii region W5 \citep{koenig08} or in the giant molecular cloud G216-2.5 \citep{megeath09}. Note that other methods, also based on \emph{Spitzer} colors and magnitudes, have been developed to find YSO candidates \citep[e.g.][]{hartmann05,harvey07,robitaille08,chavarria08}.

\subsection{Final census of YSO candidates}
\label{subsec:YSO_census}

We finally find 856~YSO candidates in \vul: 239~are likely protostars with infalling envelops (class$\;$0/I), 464~are disk-bearing stars (class$\;$II), and 153~are class$\;$III objects with very little circumstellar material. Among the class$\;$I objects, 15~YSO candidates are likely deeply embedded protostars, and 89~have a spectral index consistent with the Flat Class. Among the class$\;$II objects, 85~are highly reddened class$\;$II YSOs that were misclassified as protostars based on IRAC criteria only. A further 21 sources are classified as likely transition disk class$\;$II objects. 

Except for a few cases, all YSO candidates have detections in all four IRAC bands, about half of them have 2MASS detections in all three JHK bands, 75\% have 24\mic~counterparts (it reaches 100\% for class$\;$0/I objects) and 8\% have 70\mic~detections. Table~\ref{tab:yso_param} gives the coordinates and fluxes of all YSO candidates. Table~\ref{tab:contamin_param} in the appendices gives the coordinates, fluxes and type of the \emph{contaminants} identified in \vul. 

To test the robustness of our analysis relative to the interstellar extinction, we have used the visual extinction map of \vul~and the formulae provided by \citet{flaherty07} to deredden IRAC fluxes. We run the exact same procedure as described above, and we find that the YSO census is only marginally different from the uncorrected case (except for the class$\;$III population that is reduced by 35\%). This justifies the choice of [5.8]-[8.0] as an extinction-free indicator for finding YSOs. 

\begin{deluxetable}{l@{\extracolsep{0.25cm}}c@{\extracolsep{0.15cm}}r@{.}@{\extracolsep{0cm}}l@{\extracolsep{0.25cm}}r@{.}@{\extracolsep{0cm}}l@{\extracolsep{0.15cm}}r@{.}@{\extracolsep{0cm}}l@{\extracolsep{0.15cm}}r@{.}@{\extracolsep{0cm}}l@{\extracolsep{0.25cm}}r@{.}@{\extracolsep{0cm}}l@{\extracolsep{0.cm}}r@{.}l@{\extracolsep{0.cm}}r@{.}@{\extracolsep{0.cm}}l@{\extracolsep{0.cm}}r@{.}@{\extracolsep{0.cm}}l@{\extracolsep{0.25cm}}r@{.}@{\extracolsep{0.cm}}l@{\extracolsep{0.cm}}r@{.}@{\extracolsep{0cm}}l@{\extracolsep{0.1cm}}c} \centering
   \tabletypesize{\footnotesize}
  \tablewidth{0pt}
   \tablecaption{List of YSO candidates and their 2MASS, IRAC and MIPS photometry. \label{tab:yso_param}}
   \tablehead{\colhead{}& \multicolumn{3}{c}{Galactic} & \multicolumn{6}{c}{2MASS} & \multicolumn{8}{c}{IRAC} & \multicolumn{4}{c}{MIPS} \\
   \cmidrule(r){2-4}  \cmidrule(r){5-10}  \cmidrule(r){11-18}  \cmidrule(r){19-22}\\[-2.2ex]
   GLIMPSE source name & Glon & \multicolumn{2}{c}{Glat} & \multicolumn{2}{c}{J} & \multicolumn{2}{c}{H} & \multicolumn{2}{c}{Ks} & \multicolumn{2}{c}{[3.6]} & \multicolumn{2}{c}{[4.5]} & \multicolumn{2}{c}{[5.8]} & \multicolumn{2}{c}{[8.0]} & \multicolumn{2}{c}{[24]} & \multicolumn{2}{c}{[70]} & Class}
   \startdata 
SSTGLMC G060.3261-00.6407 & 60.3261 &     -0&6407 & .&. & .&. & .&. &       8&68 &       8&19 &       7&74 &       7&07 &       4&55 & .&. & II\\
SSTGLMC G060.1815-00.5632 & 60.1816 &     -0&5632 &       12&71 &       10&61 &       9&45 &       8&31 &       8&22 &       7&89 &       7&42 &       7&16 & .&. & III\\
SSTGLMC G060.3304-00.6042 & 60.3304 &     -0&6042 & .&. & .&. & .&. &       12&99 &       11&40 &       10&71 &       10&14 &       4&78 &      -5&65 & 0/I\\
SSTGLMC G060.3331-00.6065 & 60.3331 &     -0&6065 &       16&04 &       14&38 &       13&41 &       11&79 &       10&83 &       10&12 &       9&09 &       5&89 & .&. & 0/I\\
SSTGLMC G060.2117-00.6700 & 60.2118 &     -0&6700 &       14&69 &       13&61 &       13&05 &       12&30 &       11&62 &       11&24 &       9&95 & .&. & .&. & II\\
SSTGLMC G060.2898-00.6427 & 60.2898 &     -0&6427 & .&. & .&. & .&. &       13&14 &       12&17 &       11&44 &       10&74 &       6&98 & .&. & 0/I\\
SSTGLMC G060.2868-00.6351 & 60.2868 &     -0&6351 & .&. & .&. & .&. &       12&90 &       11&92 &       11&17 &       10&32 &       7&65 & .&. & 0/I\\
SSTGLMC G060.1319-00.5355 & 60.1320 &     -0&5355 &       10&84 &       8&99 &       8&01 &       7&30 &       7&24 &       6&94 &       6&70 &       5&44 & .&. & III\\
SSTGLMC G060.2180-00.6362 & 60.2180 &     -0&6362 &       10&68 &       8&79 &       7&86 &       7&24 &       7&28 &       6&85 &       6&55 &       5&27 & .&. & III\\
SSTGLMC G060.1975-00.6052 & 60.1975 &     -0&6052 & .&. & .&. & .&. &       13&63 &       12&32 &       11&28 &       10&38 &       7&24 & .&. & 0/I\\
 \enddata
\tablecomments{Galactic coordinates are in units of degrees (\degr). Table~\ref{tab:yso_param} is published in its entirety in the electronics edition of the \emph{Astrophysical Journal}. A portion is shown here for guidance regarding its form and content.}
\end{deluxetable}

\subsection{Reliability and completeness}
\label{subsec:completeness}

In the present study, we make the deliberate choice to favor the highest possible detection reliability, at the expense of a modest completeness figure. This choice was largely motivated by the need to automate the search for YSOs over such a large area of the sky. 

As mentioned in Section~\ref{subsec:initial_cat}, we achieve a detection reliability better than 98\% for the set of IRAC magnitude cuts chosen for populating our initial catalog. However the actual detection reliability depends to a certain extent on the level of the background diffuse emission. 
An adequate assessment of the completeness of our YSOs catalog would require to carry out a large spectroscopic survey of the red objects in \vul~in order to identify genuine YSOs, and then compare the results to our catalog. This is impractical though, given the large number of (faint) objects in \vul. 

Considering the stringent criteria used to reach a reliability over 98\%, we expect a noticeable number of YSOs to be excluded from our initial point source catalog. For instance, any saturated sources, that are either bright sources on faint background or faint sources on bright background, are not considered in our analysis. There are very few saturated data in IRAC images so that the YSOs selection is basically unaffected by this effect. Still, we expect that a fraction of YSO candidates will not have 24\mic~counterparts because of saturation issues (especially around Sh2-88B and Sh2-87 where the background is high). Extended sources are also excluded based on the selection procedure used in this work (see Section~\ref{sec:YSO}), but their actual number is expected to be fairly small at 2.3~kpc. Point sources with too large a magnitude error are not considered either. We set the limit for bad photometry to be $\sigma_{\rm mag}>0.2$~mag. IRAC bands 3 and 4 being the noisiest bands, the selection criterion on $\sigma_{\rm mag}$ mostly impacts [5.6] and [8.0]~detections.

The completeness of our YSO catalog is also dependent on the evolutionary stage of the object. For instance, the wavelength band 2-24\mic~is well suited for detecting and identifying disks around young stars (class$\;$II), but longer wavelengths are more appropriate for observing early-stage objects (class$\;$0/I). We believe that our sample of IRAC-selected YSO candidates may lack a significant fraction of these protostars. 
For instance, we looked for a possible correlation between the spectral index $\alpha_{\rm IRAC}$ and the visual extinction of the YSO candidates expecting to find a population of protostars with high values of $\alpha_{\rm IRAC}$, i.e. the youngest objects, to be the most extinct/embedded objects. However we did not  find such correlation in our sample which suggests that we might be missing the most embedded phases of star formation based on IRAC and MIPSGAL~24\mic~data only. Similarly, the population of class$\;$III YSOs which lacks near-IR excess emission is strongly underestimated in our study, even if they are genuine YSOs possessing H$\alpha$ or X-ray emission. For all these reasons, it is very difficult to assess the completeness of our YSO catalog with confidence. 
We use the evolutionary models of \citet{baraffe98} to compute the mass of a 2~Myr old star at 2.3~kpc with no IR-excess emission having a 3.6\mic~magnitude of 14~mag; and we find a mass of $\sim$1~M$_\sun$. Considering that the 8\mic~channel of IRAC is the less sensitive channel (limit magnitude of 10.8$\;$mag), and that most YSOs have significant IR-excess emission, we estimate our YSO catalog to be complete down to a few solar masses.

Lastly, we have to account for the possible misidentification of YSO candidates with evolved stars. The circumstellar environment of Planetary Nebulae (PNe) or Asymptotic Giant Branch (AGB) stars are rich in warm dust grains emitting in the near- mid-IR such that these objects occupy the same locus in color-color diagrams as YSOs \citep{hora08,srinivasan09}. Following the analysis of \citet{robitaille08}, we estimate the fraction of evolved stars in our catalog of YSO candidates to be about 25\%, of which most are AGB stars. We will argue in Section~\ref{subsec:clustering} that AGB stars could be further excluded from the sample of YSO candidates based on clustering criteria.

\section{Results and discussion}
\label{sec:result}

\subsection{Star Formation Efficiency}
\label{subsec:sfe}

We estimate the current star formation efficiency in Vulpecula by comparing the mass of the gas reservoir, M$_{\rm cloud}$, with the mass that has turned into stars during the last few million years, M$_{\rm YSO}$. The Star Formation Efficiency (SFE) is then derived as follows: 
\begin{equation}
\rm{SFE}=\frac{M_{\rm YSO}}{M_{\rm YSO}+M_{\rm cloud}}
\end{equation}
We compute the mass of the cloud using our extinction map of Vulpecula and the formula relating the column density and the visual extinction N$_{\rm H}$/A$_{\rm V}=1.37\times10^{21}$~cm$^{-2}$.mag$^{-1}$ assuming a value of $R_{\rm V}=5.5$ typical for molecular clouds as suggested in \citet{evans09}. We use a contour of A$_{\rm V}=7\,$mag to delineate the active star forming region on the A$_{\rm V}$ map of Vulpecula. We find an average mass surface density of 16.2~M$_\sun$.pc$^{-2}$. We integrate the surface density over the $2.7\times 10^3$~pc$^2$ enclosing the A$_{\rm V}>7$ region, and we find a cloud mass $M_{\rm cloud} = 4.5\times 10^4$~M$_\sun$. 

Since our YSO sample represents only the high-end of the Initial Mass Function (IMF), as we estimated in Section~\ref{subsec:completeness}, we need to account for the missing mass of the total YSO population in order to derive a relevant M$_{\rm YSO}$. We use the IMF of \citet{kroupa01} to compute the number of stars with masses ranging from 0.01 to 50~M$_\sun$. We normalize the IMF assuming 510~stars have masses $>$1~M$_\sun$ within the A$_{\rm V}>7$ region, and we find that \vul~should contain $2\times 10^4$ YSOs with a total mass of M$_{\rm YSO}=4200$~M$_\sun$. We thus obtain a SFE of $\sim$8\%, similar to other star forming regions in nearby molecular clouds \citep{evans09}. 

\subsection{Spatial distribution of YSO candidates}
\label{subsec:clustering}

\begin{landscape}
\begin{figure}\centering
    \includegraphics[width=1.3\textwidth,angle=0]{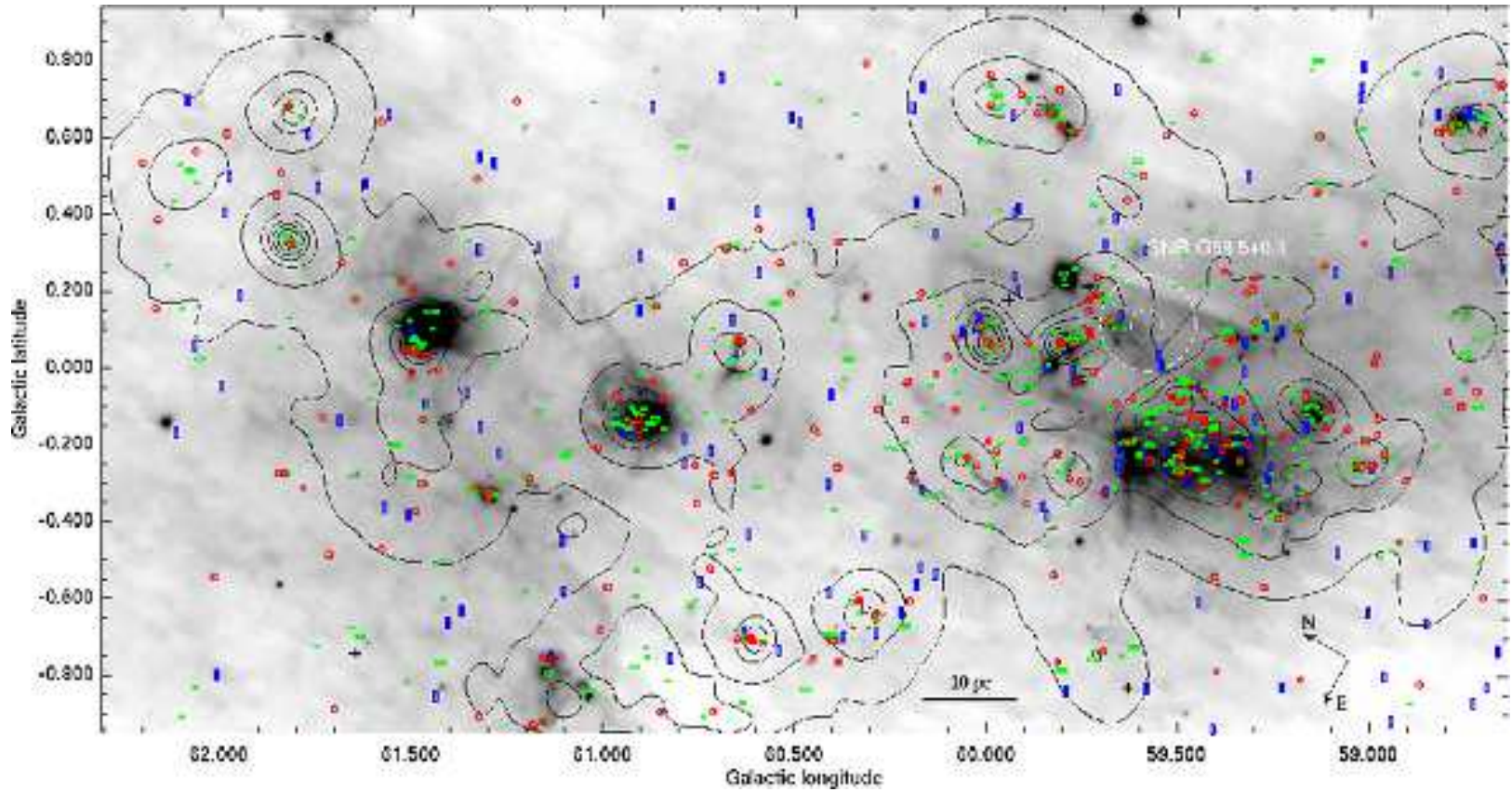}
    \caption{Distribution of YSO candidates in \vul~displayed over the 70\mic~negative image from MIPSGAL. Red circles represent envelops/Class$\;$I YSO candidates, green horizontal rectangles represents disks/Class$\;$II objects, and blue vertical rectangles are optically thin disks/Class$\;$III objects. Black contours represent the surface density of YSO candidates in \vul~as described in the text. Their distribution is well correlated with the mid-IR morphology. They are found at the tip of most pillars and around \hii regions. The large and small dashed circles around $(l,b)=(59.58,0.12)$ indicate the diameter of the supernova remnant SNR~G59.5+0.1 as published in \cite{taylor92} and \cite{green06} respectively. The orientation and scale of the image are also indicated.  \label{fig:M70_yso_snr}}
\end{figure}
\end{landscape}

The spatial distribution of YSO candidates in \vul~is presented in Figure~\ref{fig:M70_yso_snr}. YSO candidates are not distributed randomly in the field. We rather find several coherent structures, or groups of YSO candidates, which in most cases are reminiscent of the mid-infrared morphology of the complex. 
For instance the highest densities of YSO candidates are located close to, or within the three \hii regions. 
Other YSOs appear to line up along InfraRed Dark Clouds (IRDC), e.g. at $(l,b)=(59.98,0.06)$ at the pedestal of the pillar \emph{VulP5} (see Figure~\ref{fig:pillars_color}); or along bright contrasted structures of the extended 24\mic~emission, e.g. at $(l,b)=(61.82,0.33)$ or $(l,b)=(59.80,0.64)$. YSO candidates have also been identified at the tip of most newly discovered pillar structures (see Table~\ref{tab:pillars_detail} and Sections~\ref{subsubsec:pillars} and~\ref{subsec:pillarSF}). 

If we assume the IMF parameters presented in Section~\ref{subsec:sfe}, we estimate that the average surface density of YSO candidates is 7.4~YSO.pc$^{-2}$. This value is consistent with the typical surface density found in other star forming regions, e.g. 13.0~YSO.pc$^{-2}$ in Serpens \citep{harvey07} or 3.3~YSO.pc$^{-2}$ in Lupus \citep{merin08}.
We build a surface density map of YSO candidates following the definition of \citet{chavarria08}. At each point of a 5\arcsec~grid, we compute the surface density of YSOs as $$\sigma=\frac{N}{\pi r_N^2},$$ where $r_N$ is the distance to the $N=5$ nearest neighbor. This image is then convolved with a gaussian of FWHM = 1\arcmin~to obtain a smooth contour map of the surface density. Figure~\ref{fig:M70_yso_snr} shows the surface density contours over a 70\mic~image of \vul. It clearly delineates the mid-IR-bright \hii regions, plus other groups of YSO candidates associated with fainter compact mid-IR sources, e.g. at $(l,b)=(60.60,-0.70)$. The surface density peaks at $\sim$50~YSO.pc$^{-2}$ at the position $(l,b)=(59.80,0.064)$ which is coincident with the BLAST source V30, also known as IRAS~19410+2336. This source is a candidate Hyper Compact \hii region according to \citet{chapin08}, and about 15 of our YSO candidates appear to be grouped around it (see Section~\ref{subsec:pillarSF}). 

In addition to this population of clustered YSO candidates, there is a population of distributed IR-excess sources. \citet{robitaille08} argue that a fraction of these isolated red objects could be AGB stars mistaken for YSO candidates in our selection process (they occupy the same locus in the C-C diagram, cf Section~\ref{sec:YSO}). However \citet{koenig08} argue that the isolated YSO candidates could be genuine YSOs that formed from isolated events or that were ejected\footnote{Assuming a relative velocity of 3~km~s$^{-1}$ between a young star and its parental cloud, a YSO could travel 6~pc in less than 2~Myr. This explains how genuine YSOs can contribute to the distributed population in star forming regions.} from their birthplace due to gravitational interactions with other members of their native star cluster. 

We now focus our analysis on the population of clustered YSO candidates since they are more reliable tracers of triggered star formation events. 
Various methods can be used to discriminate between distributed/clustered populations. For instance the two-point correlation function was used by \citet{karr03} on W5 and by \citet{indebetouw07} on M16, while the minimum spanning tree technique was used on W5 by \citet{koenig08} and on NGC~1333 by \citet{gutermuth08}. Both methods provide valuable qualitative results but dynamical measurements are absolutely necessary to assign a definitive membership to a given YSO. Since such data are not available for \vul~we opt for the simplest filtering, namely the nearest neighbor, to gain insight into the clustering properties of our YSO sample. In practice, we consider all YSO candidates and keep only those that have their nearest neighbor YSO within 3\arcmin~(2~pc at 2.3~kpc). The result of this selection is a much more clustered distribution of YSO candidates that traces mid-IR morphology even more closely (192 YSO candidates are marked as isolated objects).

\begin{figure} \centering
  \begin{tabular}{l}
    \includegraphics[width=0.45\textwidth,angle=0]{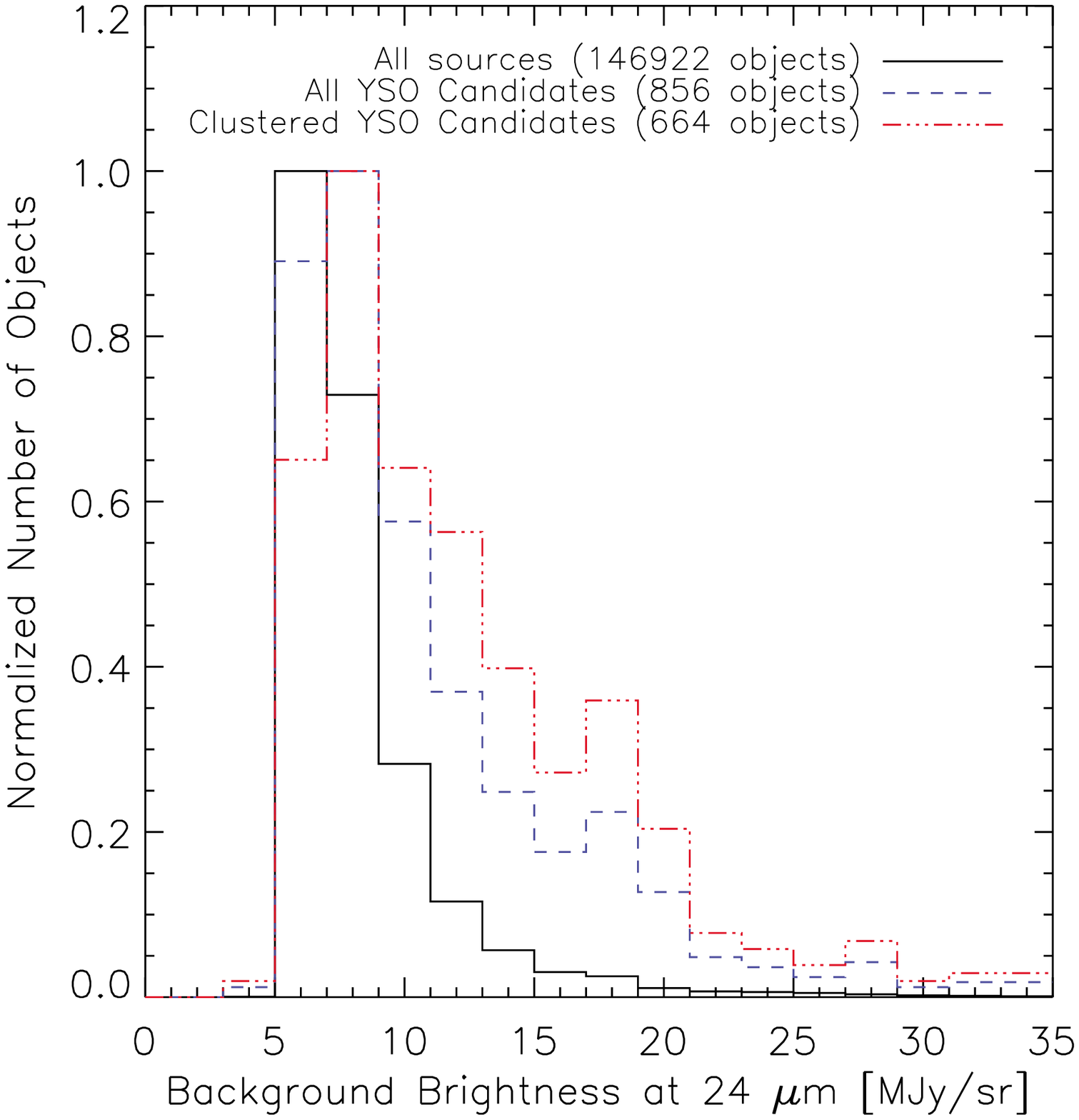} \\
    \includegraphics[width=0.45\textwidth,angle=0]{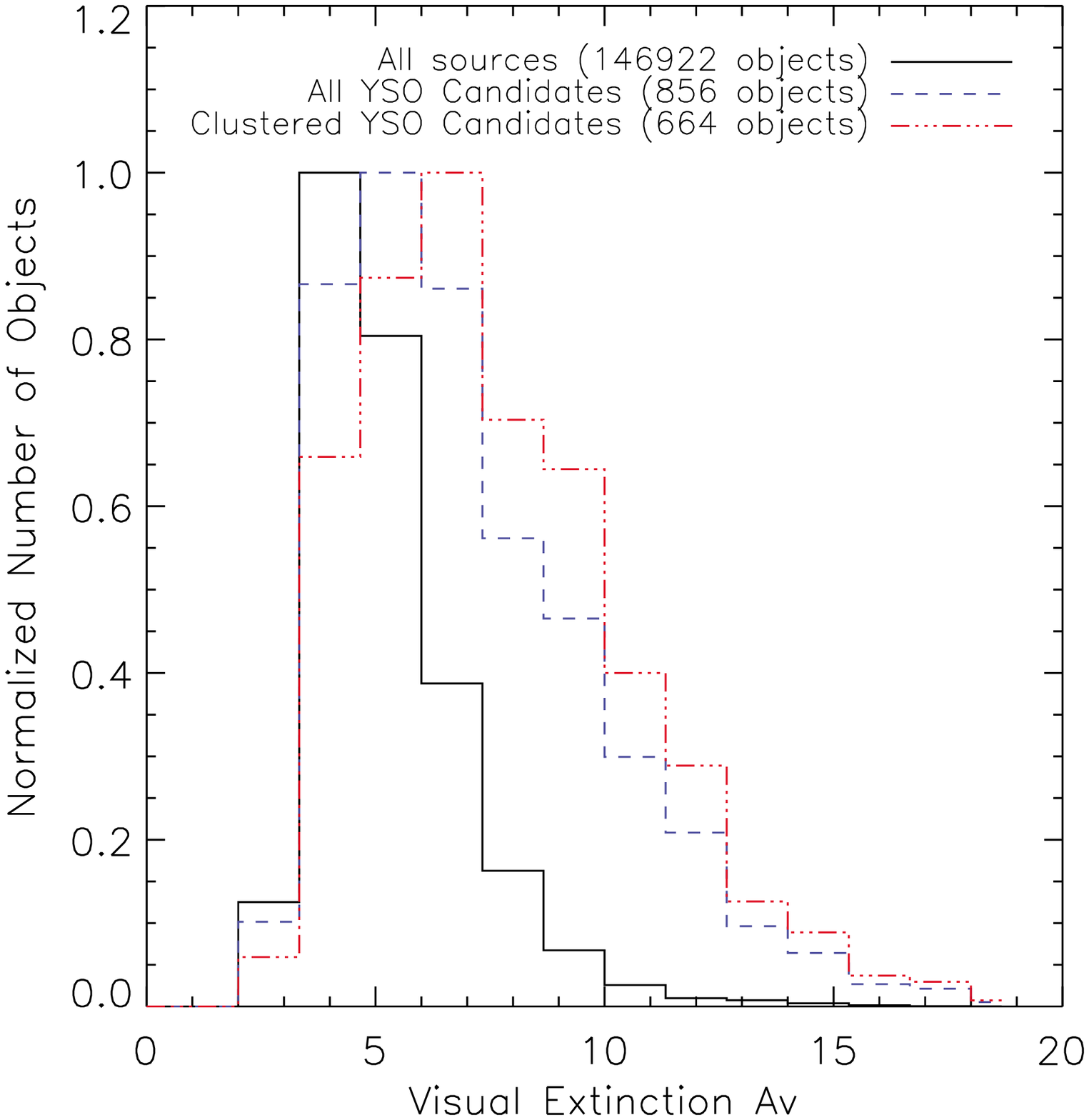}
  \end{tabular}
    \caption{Histograms of 24\mic~background brightness (\emph{upper panel}) and visual extinction (\emph{lower panel}) associated with each point sources of our initial catalog. YSO candidates appear to sit preferentially on high 24\mic~background as well as in high extinction, i.e. embedded, regions. This is even more pronounced for clustered YSO candidates.   (A color version of this figure is available in the online journal.) }
    \label{fig:yso_av_correl}
    \end{figure}

We also look for trends in the environment of YSO candidates compared to field stars. The top panel of Figure~\ref{fig:yso_av_correl} shows that YSO candidates sit preferentially on regions of bright 24\mic~background emission compared to the more evolved stars. Since the 24\mic~emission is mostly due to the thermal emission of small dust grains excited by UV radiation, the trend for YSO candidates to sit on bright mid-IR background implies that the identified YSO candidates are largely associated with photon dominated regions (PDR) and \hii regions. This is consistent with known scenarios of triggered star formation mechanisms and with previous studies of star forming regions \citep[e.g.][]{koenig08}.
We also compare the visual extinction associated with the YSO population to the entire population of point sources found in our initial catalog (see bottom panel of Figure~\ref{fig:yso_av_correl}). We find that YSO candidates are mostly located in regions of high extinction compared to field stars. This is again consistent with the idea that infant stars still live in the dense and dusty cloud from which they were born. 
Remarkably the environmental trends mentioned above are even more pronounced for the population of clustered YSO candidates (see Figure~\ref{fig:yso_av_correl}). 

Note that \citet{chapin08} identified a few BLAST sources in the direction of \vul~that are actually located at 8.5~kpc from the Sun (radial velocity around $-5$~km~s$^{-1}$), i.e. in the Perseus arm. A dozen of our YSO candidates seem to  be associated with these distant BLAST sources so that these might belong to the Perseus arm.

\subsection{Spectral Energy Distribution}
\label{subsec:sed}

\begin{figure*}\centering
    \includegraphics[width=1.\textwidth,angle=0]{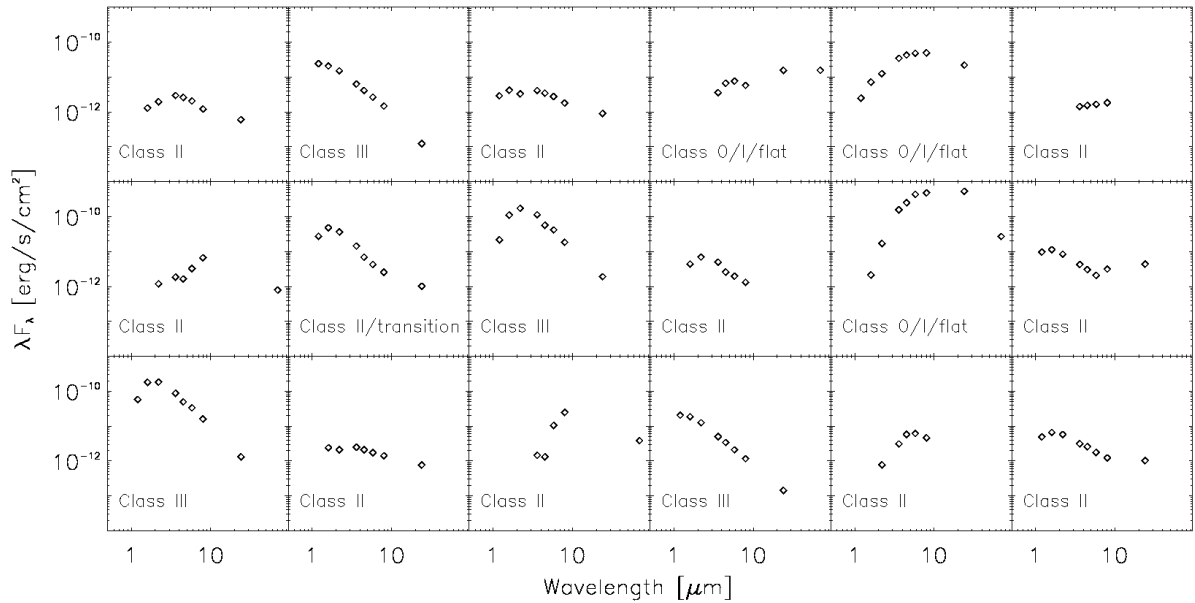}
    \caption{Sample of SEDs from our catalog of YSO candidates. Almost all YSO candidates have detection in all four IRAC bands, 75\% have detections at 24\mic, 6\% at 70\mic~and about half in J, H and K bands. Our classification of each YSO candidates is shown on individual plots. The axis scale is the same for each subplot.
    \label{fig:SED_example}}
\end{figure*}

We have built Spectral Energy Distributions (SEDs) for all of our YSO candidates. Figure~\ref{fig:SED_example} presents a small SED sample with the Class designation associated with each object. 

We used the SED fitter tool developed by \citet{robitaille07} to extract the physical parameters of our YSO candidates. However, the limited wavelength coverage (near- to mid-IR) resulted in strong degeneracies over the output parameters, and we were unable to gain reliable information on the physical properties of our YSO candidates.

We looked for possible associations in the BLAST field to extend the wavelength coverage to the sub-millimeter, but in most cases several YSO candidates fell within the BLAST beam rendering the association ambiguous. For the YSO candidates that could be uniquely associated with a BLAST source, \citet{chapin08} provide physical parameters derived from SED fitting. They find clump masses ranging from 40 to 400~M$_{\sun}$ and temperatures from 19 to 28~K. We also use \citet{molinari08a} diagnostic diagrams based on MIPS [24-70] and BLAST [250-500] colors to identify the most massive YSOs in \vul. \citeauthor{molinari08a} show that the high-mass analogues of the low- or intermediate-mass Class$\;$0 objects have distinctive MIPS colors. From our sample of YSO candidates, we find 6 objects with $[24-70] >3$, which is indicative of an SED peaking longwards of 70\mic~presumably due to a massive infalling envelop. Three of these red MIPS sources are located next to the brightest BLAST source V30 and are barely discernible on the map, two have no BLAST counterparts, and one might have a BLAST counterpart but the association is ambiguous as another YSO candidate falls within the BLAST beam. 

Finally we searched the IRSA for Bolocam sources, and we found 24 sources in \vul. Among them we find isolated sources not associated with any YSO candidates, these could be YSOs in their earliest evolutionary phases, i.e. starless cores emitting in the millimeter only. The other Bolocam sources are usually located close to the peaks of the YSO surface density and are associated with groups of YSO candidates. Note that almost all Bolocam sources have BLAST counterparts over the Vulpecula BLAST field.

\subsection{Cases of triggered star formation?}
\label{subsec:trigger}

\subsubsection{The case of SNR~G59.5+0.1}
\label{subsec:snr}

Shock waves generated by supernovae can potentially trigger episodes of star formation at the interface between the SNR and the ISM \citep{melioli06}. The supernova remnant SNR~G59.5+0.1 was first detected by \cite{taylor92} in the direction of \vul~at the position $(l,b)=(59.58,0.12)$. \citeauthor{taylor92} describe this object as a shell-type SNR with non-thermal spectral index ($\alpha<-0.4$) and a diameter of 15\arcmin. They also mention the close proximity and possible association of the SNR with the \hii region Sh2-86. We investigate the possibility of such an association and potential signs of triggered star formation.

\cite{guseinov03} computed the distance to SNR~G59.5+0.1 using an empirical formula that relates the surface brightness of a SNR ($\Sigma$) with its diameter ($D$). They calibrated the $\Sigma-D$ relation against SNRs of known distances, and they found that SNR~G59.5+0.1 lies at $\sim$11~kpc from the Sun. This would imply that SNR~G59.5+0.1 is not associated with the \hii region Sh2-86. We stress however that \citeauthor{guseinov03} computed the distance to G59.5+0.1 based on data from the SNR catalog of \cite{green06} ($\Sigma_{1GHz}=18.1\times10^{-21}$~W.m$^{-2}$.Hz$^{-1}$.sr$^{-1}$ and $D=5$\arcmin~from observations at 1720~MHz), whereas \citeauthor{taylor92} report different values from observations at 327 and 4850~MHz ($\Sigma_{1GHz}=0.7\times10^{-21}$~W.m$^{-2}$.Hz$^{-1}$.sr$^{-1}$ and $D=15$\arcmin). 
We use the VGPS 21-cm data to look for traces of the SNR and settle this discrepancy. We find a circular structure centered at the position of the SNR with a diameter of 15\arcmin, which confirms the value found by \citeauthor{taylor92}. Therefore we compute the distance to SNR~G59.5+0.1 using \citeauthor{guseinov03}'s formula with \citeauthor{taylor92}'s measurements instead of \citeauthor{green06}'s; and we find that the SNR likely lies between 2.1~kpc and 5.3~kpc from the Sun provided the given uncertainties of the $\Sigma-D$ relation. This distance estimate reconciles the hypothesis that SNR~G59.5+0.1 is indeed associated with Sh2-86.

\citet{reach06} have searched for infrared counterparts to the known SNRs in the inner galactic plane using IRAC observations from the GLIMPSE survey, however they did not detect SNR~G59.5+0.1 because of the high confusion level in the vicinity of Sh2-86 structured extended emission. For the same reason, we could not detect the SNR on MIPSGAL~24\mic~images either. No morphological clues were found in the ancillary data mentioned in Section~\ref{subsec:anci}.
Nevertheless, the distribution of YSO candidates around SNR~G59.5+0.1 reveals two groups of YSOs apparently lining up along two arcs at the northern and southern rims of SNR~G59.5+0.1 (cf Figure~\ref{fig:M70_yso_snr}). Assuming that the actual diameter of the SNR is indeed 15\arcmin, the two groups do not lie exactly on the rim of the shell but slightly outside. We still believe that these YSO overdensities could have arisen from the interaction of the expanding shell of SNR~G59.5+0.1 with the neutral gas of Sh2-86.
According to \citet{xu05}, the age of a shell-type SNR such as SNR~G59.5+0.1 with a diameter of 10~pc (assuming a 15\arcmin~diameter and a distance of 2.3~kpc) ranges from $10^3$ to $10^4$ years old. This timescale is comparable to the lifetime of the first phases of star formation, making the SNR a possible trigger for the surrounding YSO candidates.
These clues tend to confirm the association of SNR G59.5+0.1 with Sh2-86, but since we could not find unequivocal evidences for the SNR-\hii region association, a fortuitous spatial coincidence cannot be excluded.

\subsubsection{Photoevaporation at the tip of the pillars}
\label{subsec:pillarSF}

An interstellar cloud exposed to the ionizing radiation of a newly formed star is compressed by an ionization-shock front which can focus the neutral gas into a compact globule. This mechanism is called radiation-driven implosion (RDI) and is described in \citet{bertoldi89}. In some cases, this leads to the birth of a second generation of stars located at the interface of the ionized and neutral gas \citep[e.g.][]{reach09}. 

Even though we could not identify the sculptor of the pillars described in Section~\ref{subsubsec:pillars}, we expect the RDI mechanism to be at work in these objects. Figure~\ref{fig:pillars_color} shows the location of our YSO candidates in pillars \emph{VulP1} to \emph{VulP6} and \emph{VulP11}. We find at least one YSO candidate at the tip of most of the pillars (see Figure~\ref{fig:pillar_24mu_ID} and last column of Table~\ref{tab:pillars_detail}). YSO candidates can also be found along the pillars when the gas is protruding and exposed to the ionizing star (e.g. YSO on the West side of \emph{VulP3} on Figure~\ref{fig:pillars_color}). BLAST also detected sources at the tip of every pillar found within the Vulpecula BLAST field (e.g. bottom panel of Figure~\ref{fig:pillars_color}). 

\subsubsection{Embedded star cluster}
\label{subsec:embedcluster}

Cr404 is a young \citep[9~Myr old, ][]{bica08} star cluster located at $(l,b)=(59.14, -0.11)$ embedded in Sh2-86. We find a high surface density of YSO candidates associated with this cluster. 11 Class$\;$II-III and 3 Class$\;$I YSO candidates surround the mid-IR peak emission of Cr404. 
BLAST observations further reveal that the embedded cluster is coincident with the bright sub-millimeter source V18, also identified as IRAS 19403+2258. \citet{chapin08} classify this source as a candidate Hyper Compact (HC) \hii region, i.e. the precursor of a high-mass star, and they derived a mass of $\sim$150~M$_\sun$ and a temperature of $\sim$30~K for this object. 
The simultaneous presence of a HC \hii region ($\sim10^5$~yr lifetime) and Class$\;$II YSOs (few $10^6$~yr lifetime) around a young star cluster such as Cr404 is consistent with the results of \citet{molinari08} which suggests that the most massive objects in a cluster are the last ones to form.

Other cases of HC \hii regions associated with groups of \emph{Spitzer}-selected YSO candidates are found in the BLAST field. For instance, the BLAST source V30, the brightest  and most massive clump according to \citet{chapin08} is coincident with the highest density of YSO candidates in the whole 6.6~square degree map at $(l,b) = (59.80,0.064)$.

\subsubsection{Large scale propagating Star formation}
\label{subsec:propage}

Based on radio observations of \hi supershells and on energetics arguments, \citet{ehlerova01} estimate that the \hi shell GS061+00+51, which encompasses Sh2-87 and Sh2-88, is 6-7 times larger and 3-4~Myr older than the shell GS59.7-0.4+44 that hosts Sh2-86. \citeauthor{ehlerova01} attribute these age and size differences to the delayed expansion of \hii regions, and they suggest that star formation might be propagating from Sh2-88 to Sh2-86. Previously, \cite{turner86} suggested that the morphology of the three \hii regions was shaped by a single supernova shock wave associated with the fossil \hii region Lynds~792 $(l,b)=(60.81, 2.24)$. Since Lynds~792 is approximately equidistant from the three \hii regions, \citeauthor{turner86}'s scenario implies that star formation would be triggered simultaneously, as opposed to sequentially in the case of \citeauthor{ehlerova01} 
We now test the validity of these scenarios using the supplemental information gathered in this work.

\begin{figure}\centering
    \begin{tabular}{c}
    \includegraphics[width=0.44\textwidth,angle=0]{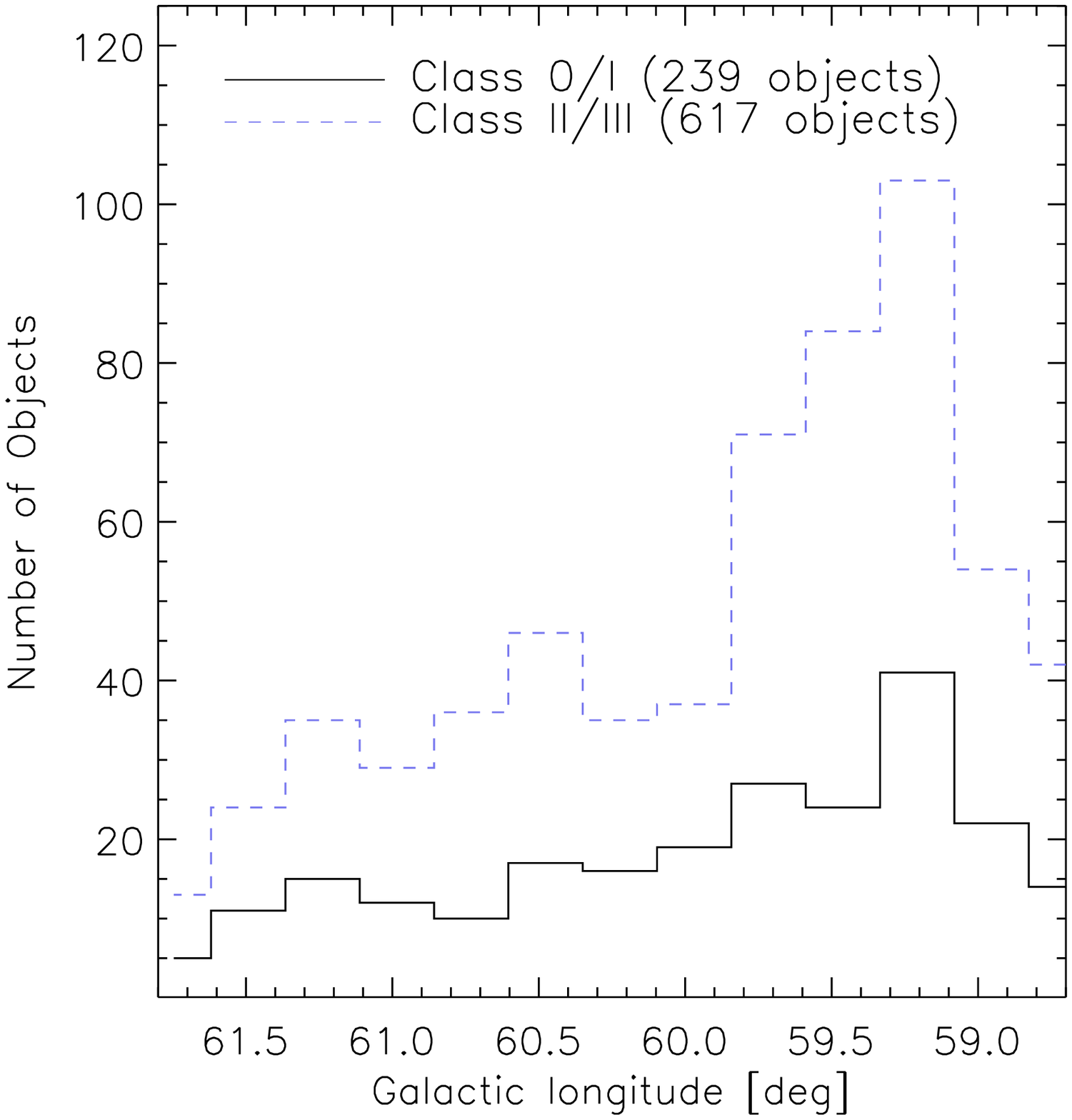} \\
    \includegraphics[width=0.44\textwidth,angle=0]{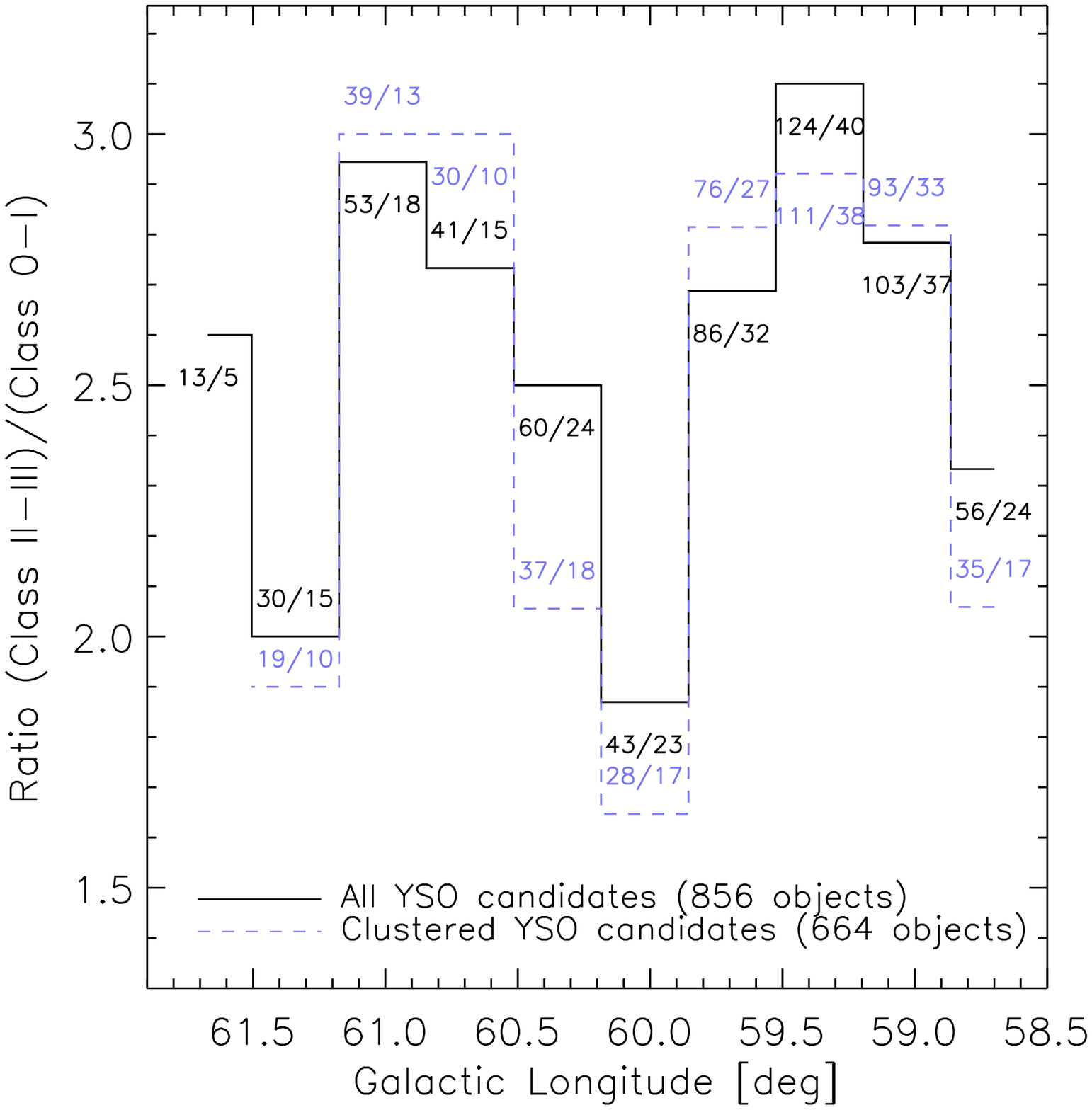}
    \end{tabular}
    \caption{\emph{Top panel}: Distribution of class$\;$0/I and~II/III candidates as a function of Galactic longitude (bin size $\sim$15\arcmin). The number of young stellar objects increases towards lower longitudes. \emph{Bottom panel}: Evolution of the class$\;$II/III to class$\;$0/I ratio as a function of longitude. The number of class$\;$0/I and~II/III in each bin is indicated on the figure.  (A color version of this figure is available in the online journal.)}
    \label{fig:yso_gradient_longitu}
    \end{figure}

Figure~\ref{fig:yso_gradient_longitu} shows the evolution of the number of YSO candidates as a function of their Galactic longitude. The increasing number of YSO candidates towards Sh2-86 seems to favor the scenario of \citeauthor{ehlerova01} since the youngest star forming region should indeed possess the largest number of young stellar objects, assuming identical IMFs and instantaneous star formation. Nevertheless the shape of this histogram could also be interpreted as Sh2-86 being a more active star forming region than the other two Sharpless objects, creating more YSOs from a larger molecular reservoir, which is unrelated to any process of propagating star formation. 
For a proper interpretation of the Figure~\ref{fig:yso_gradient_longitu} histogram, we need to compare the typical lifetime of a YSO and the timescale necessary for star formation to propagate in a molecular cloud.

\citet{nomura01} have studied self-propagating star formation using numerical simulations and found that the time delay of sequential star formation sites against the original one, $\Delta t$, correlates with their physical separation, $\Delta x$, as $\Delta t \sim 50 Myr[\Delta x/(0.5 kpc)]^{0.5}$. \cite{efremov98} have derived a similar expression based on observations  of star clusters in the Large Magellanic Cloud ($\Delta t \sim 26 Myr[\Delta x/(0.5 kpc)]^{0.4}$). Assuming Sh2-88 and Sh2-86 are separated by $\sim$80~pc, we find that star formation would take $\sim$10-20~Myr to propagate across \vul. 
Besides, \citet{evans09} estimate the lifetime of a YSO to be of the order of 1-3~Myr based on a large statistical sample of YSOs identified as part of the c2d legacy survey\footnote{http://irsa.ipac.caltech.edu/data/SPITZER/C2D}. 
If star formation was indeed propagating along the Galactic plane at the pace derived above, YSOs would only be found in a narrow slab of longitudes since they would age, and fade away in the IR, faster than the triggering shock needs to cross the molecular complex. We can already rule out this scenario based on the presence of YSOs in all three \hii regions of \vul~(see Figure~\ref{fig:M70_yso_snr}). 
It is possible however that star formation is propagating faster than previously estimated due to the pronounced inhomogeneities of the propagation medium (cf distribution of CO in Figure~\ref{fig:Av_CO}). If the propagation timescale was of the same order of the YSOs lifetime, then YSOs would be found all across \vul~with a gradient in the evolutionary phases of the YSO population as a function of longitude. 

We use the ratio of Class$\;$II-III to Class$\;$0-I as an indicator of the aging of the YSO population. Figure~\ref{fig:yso_gradient_longitu} shows that this ratio does not present a gradient with respect to longitudes. This indicates that if star formation was once triggered in \vul, then the triggering agent would have been independent of Galactic longitude, and this definitely rules out the suggestion of \citeauthor{ehlerova01} 
Nonetheless, the histogram of Figure~\ref{fig:yso_gradient_longitu} is still consistent with \citeauthor{turner86}'s scenario in which star formation occurs as an instantaneous burst in the three \hii regions.

\section{Conclusions}
\label{sec:conclu}

We have presented a thorough description of the Vulpecula OB association. We have complemented the existing observations of isolated objects in \vul~with a global view of the whole star forming complex from an infrared perspective. We exploited \emph{Spitzer} legacy surveys MIPSGAL and GLIMPSE data to identify 856 young stellar objects with IR-excess emission. We rely on the nature and properties of these objects to highlight the recent activity of star formation in the complex, and we look for evidences of triggered star formation. 

We find two populations of YSO candidates: one population of distributed objects that likely contains IR-bright evolved stars and some genuine YSOs born in isolation or ejected from their parental cloud; and another population of clustered YSO candidates whose spatial distribution correlates very well with the mid-IR morphology of the complex. YSO candidates surface density peaks locally around the three Sharpless objects, the Hyper Compact \hii regions, and other embedded star clusters like Cr404. The vigorous star forming activity around these energetic sources is consistent with scenarios of triggered star formation mechanisms. Still we cannot ascertain that these stars were born as a direct consequence of their extreme environment, nor that they would have formed in the absence of what we consider the triggering agents.
Nevertheless our analysis allowed to rule out the scenario of \citet{ehlerova01} according to which star formation was propagating from Sh2-88 to Sh2-86. We rather find that the evolutionary stage of the YSO population across \vul~is homogeneous, and thus consistent with the scenario of \citet{turner86}.

We have further reported the discovery of a dozen pillar-like structures in \vul, and we comment on their morphology from the near-IR to the radio regime. We were not able to identify the energetic source(s) responsible for the molding of the pillars, but we argue that these objects are indeed associated with the OB~association. Our finding of YSO candidates at the tip of most of the pillars is consistent with mechanisms of triggered star formation on small scales.

\acknowledgments

The authors would like to thank S. Bontemps for providing the extinction map of \vul.
This work is based on observations made with the \emph{Spitzer Space Telescope}, which is operated by the Jet Propulsion Laboratory, California Institute of Technology under a contract with NASA.
This research used the facilities of the Canadian Astronomy Data Centre operated by the National Research Council of Canada with the support of the Canadian Space Agency.
The National Radio Astronomy Observatory is a facility of the National Science Foundation operated under cooperative agreement by Associated Universities, Inc.
The Virginia Tech Spectral-Line Survey (VTSS) is supported by the National Science Foundation.
This research has made use of the SIMBAD database, operated at CDS, Strasbourg, France.
This research has made use of the NASA/ IPAC Infrared Science Archive, which is operated by the Jet Propulsion Laboratory, California Institute of Technology, under contract with the National Aeronautics and Space Administration.

{\it Facilities:} \facility{Spitzer Space Telescope}

\appendix

\section{Appendix material}

\begin{deluxetable}{l@{\extracolsep{25pt}}c@{\extracolsep{10pt}}c@{\extracolsep{15pt}}c@{\extracolsep{10pt}}c@{\extracolsep{15pt}}r@{\extracolsep{5pt}}@{ $ \times $}l@{\extracolsep{25pt}}c} \centering
\tablecolumns{8}
\tabletypesize{\small}
\tablewidth{0pt}
\tablecaption{Pillars identification, coordinates and approximate size. \label{tab:pillars_coord}}
\tablehead{\colhead{}& \multicolumn{2}{c}{Galactic}  & \multicolumn{2}{c}{Equatorial} &  \multicolumn{3}{c}{}\\
			 \cmidrule(r){2-3}  \cmidrule(r){4-5} \\[-1.8ex]
  		Pillar ID & l & b & RA$\,$(J2000) & Dec$\,$(J2000) & \multicolumn{2}{c}{Size} &  Position angle \\	 [-2.2ex]}
	\startdata
	\emph{VulP1} & 60.18& -0.31 & 296.36& 23.88 & 5-8 & 0.5 & 250 \\ 
	\emph{VulP2} & 60.03& -0.35 & 296.32& 23.74 & 1 & 0.5 & 230 \\ 
	\emph{VulP3} & 59.97& -0.31 & 296.25& 23.70 & 8 & 2 & 260  \\ 
	\emph{VulP4} & 59.81& -0.29 & 296.15& 23.58 & 3 & 1 & 268 \\ 
	\emph{VulP5} & 60.06& 0.10 & 295.91& 23.99 & 4 & 2 & 335 \\ 
	\emph{VulP6} & 60.11& 0.21 & 295.83& 24.10 & 5 & 2-4 & 350 \\ 
	\emph{VulP7} & 60.38& 0.21 & 295.98& 24.33 & 9 & 1-4 & 17 \\ 
	\emph{VulP8} & 60.90& 0.22 & 296.26& 24.78 & 11 & 4 & 55 \\ 
	\emph{VulP9} & 61.10& -0.45 & 297.00& 24.61 & 4 & 1 & 158 \\ 
	\emph{VulP10} & 60.77& -0.63 & 297.00& 24.24 & 4 & 2 & 190 \\ 
	\emph{VulP11} & 60.21& -0.45 & 296.52& 23.84 & 4 & 2 & 280 \\ 
	\emph{VulP12} & 59.60& -0.14 & 295.90& 23.47 & 3 & 1 & 133 \\ 
	\emph{VulP13} & 59.51& -0.22 & 295.92& 23.36 & 3 & 0.5 & 160  \\ 
	\emph{VulP14} & 59.49& -0.19 & 295.88& 23.35 & 2 & 0.5 & 150 \\ 
	\enddata
\tablecomments{Coordinates and position angles are in units of degrees. Pillar sizes are in units of arcminutes. Position angles are given between the long axis of the pillar and the West axis, positive values are towards the North.}
\end{deluxetable}

\begin{deluxetable}{l@{\extracolsep{20pt}}c@{\extracolsep{10pt}}c@{\extracolsep{20pt}}c@{\extracolsep{10pt}}c@{\extracolsep{10pt}}c@{\extracolsep{10pt}}c@{\extracolsep{10pt}}c} \centering
\tablecolumns{8}
\tabletypesize{\footnotesize}
\tablewidth{0pt}
\tablecaption{Morphological description of the pillar structures identified in \vul. \label{tab:pillars_detail}}
\tablehead{\colhead{}& \multicolumn{2}{c}{MIPS} & \multicolumn{4}{c}{IRAC}&\colhead{} \\
			 \cmidrule(r){2-3}  \cmidrule(r){4-7} \\[-1.8ex]
  Pillar ID & [70] & [24] & [8.0] & [5.8] & [4.5] & [3.6] & YSO\tablenotemark{a}\\  [-2.2ex]}
\startdata
			 & Bright~Pillar & & & & & &  \\ [-1ex]
			 \raisebox{1.5ex}{\emph{VulP1}} & Compact Source & \raisebox{1.5ex}{Structured} & \raisebox{1.5ex}{Structured} & \raisebox{1.5ex}{Faint} & \raisebox{1.5ex}{None} & \raisebox{1.5ex}{Faint} & \raisebox{1.5ex}{Y} \\ [1ex]
			\emph{VulP2} & Faint & Faint & Faint & None & None & None & N \\ [1ex] 
			 & Bright~Pillar & Bright & Bright & & & &  \\ [-1ex]
			 \raisebox{1.5ex}{\emph{VulP3}} & Compact Source & Structured & Structured & \raisebox{1.5ex}{Faint} & \raisebox{1.5ex}{Faint} & \raisebox{1.5ex}{Faint} & \raisebox{1.5ex}{Y} \\ [1ex]
			\emph{VulP4} & Faint Tip & Faint & Faint & None & None & None & Y \\ [1ex]
			 & Faint~Pillar  & Bright~Pillar & Bright~Pillar & No~Pillar & No~Pillar & No~Pillar &  \\ [-1ex]
			 \raisebox{1.5ex}{\emph{VulP5}} & Pedestal & Pedestal & Pedestal & Faint Pedestal & Faint Pedestal & Faint Pedestal & \raisebox{1.5ex}{Y} \\ [1ex]
			\emph{VulP6} & Faint & Faint & Faint & None & None & None & Y \\ [1ex]
				 & Faint & Bright & Faint & & & &  \\ [-1ex]
			 \raisebox{1.5ex}{\emph{VulP7}} & Structured & Structured & Structured & \raisebox{1.5ex}{None} & \raisebox{1.5ex}{None} & \raisebox{1.5ex}{None} & \raisebox{1.5ex}{Y} \\ [1ex]
			 & & Faint & Faint & & & &  \\ [-1ex]
			 \raisebox{1.5ex}{\emph{VulP8}} & \raisebox{1.5ex}{Faint~Tip} & Structured & Structured & \raisebox{1.5ex}{None} & \raisebox{1.5ex}{None} & \raisebox{1.5ex}{None} & \raisebox{1.5ex}{Y} \\ [1ex]
			\emph{VulP9} & Faint & Bright tip & Bow Shock & Faint Tip & None & Faint & N \\ [1ex]
			 & & Bright~Tip & & & & &  \\ [-1ex]
			 \raisebox{1.5ex}{\emph{VulP10}} & \raisebox{1.5ex}{Faint} & Possible Jets & \raisebox{1.5ex}{Faint} & \raisebox{1.5ex}{None} & \raisebox{1.5ex}{None} & \raisebox{1.5ex}{None} & \raisebox{1.5ex}{Y} \\ [1ex]
			\emph{VulP11} & Bright Tip & Bright Tip & Bright Tip & Faint Tip & Faint Tip & Faint Tip & N \\ [1ex]
			 & Bright~Tip & Bright~Tip & Bright~Tip & & & &  \\ [-1ex]
			 \raisebox{1.5ex}{\emph{VulP12}} & Pedestal & Pedestal & Pedestal & \raisebox{1.5ex}{Faint} & \raisebox{1.5ex}{Faint} & \raisebox{1.5ex}{Faint} & \raisebox{1.5ex}{Y} \\ [1ex]
			\emph{VulP13} & Faint & Bright Edges & Bright Edges & None & None & None & Y \\ [1ex]
			 & & & Bow~Shock & & & &  \\ [-1ex]
			 \raisebox{1.5ex}{\emph{VulP14}} & \raisebox{1.5ex}{Bright~Tip} & \raisebox{1.5ex}{Bright~Tip} & Bright~Tip & \raisebox{1.5ex}{Bow~Shock} & \raisebox{1.5ex}{Bow~Shock} & \raisebox{1.5ex}{Bow~Shock} & \raisebox{1.5ex}{Y} \\ [1ex]
\enddata
\tablenotetext{a}{Indicates the presence of a YSO candidate at the tip of the pillar.}
\end{deluxetable}

\begin{deluxetable}{l@{\extracolsep{0.3cm}}c@{\extracolsep{0.15cm}}r@{.}@{\extracolsep{0.cm}}l@{\extracolsep{0.3cm}}r@{.}@{\extracolsep{0.cm}}l@{\extracolsep{0.15cm}}r@{.}@{\extracolsep{0.cm}}l@{\extracolsep{0.15cm}}r@{.}@{\extracolsep{0.cm}}l@{\extracolsep{0.3cm}}r@{.}@{\extracolsep{0.cm}}l@{\extracolsep{0.cm}}r@{.}l@{\extracolsep{0.cm}}r@{.}@{\extracolsep{0.cm}}l@{\extracolsep{0.cm}}r@{.}@{\extracolsep{0.cm}}l@{\extracolsep{0.3cm}}r@{.}@{\extracolsep{0.cm}}l@{\extracolsep{0.cm}}r@{.}@{\extracolsep{0cm}}l@{\extracolsep{0.15cm}}c}  \centering
	\tablecaption{List of point sources excluded from the YSO catalog and identified as probable contaminants based on their IRAC fluxes.\label{tab:contamin_param}}
	\tabletypesize{\scriptsize}
	\tablewidth{0pt}
	\tablehead{\colhead{}& \multicolumn{3}{c}{Galactic} & \multicolumn{6}{c}{2MASS} & \multicolumn{8}{c}{IRAC} & \multicolumn{4}{c}{MIPS} \\
	\cmidrule(r){2-4}  \cmidrule(r){5-10}  \cmidrule(r){11-18}  \cmidrule(r){19-22}
	GLIMPSE source name & Glon & \multicolumn{2}{c}{Glat} & \multicolumn{2}{c}{J} & \multicolumn{2}{c}{H} & \multicolumn{2}{c}{Ks} & \multicolumn{2}{c}{[3.6]} & \multicolumn{2}{c}{[4.5]} & \multicolumn{2}{c}{[5.8]} & \multicolumn{2}{c}{[8.0]} & \multicolumn{2}{c}{[24]} & \multicolumn{2}{c}{[70]} & Type}  
	\startdata
	SSTGLMC G059.6291-00.8301 & 59.6291 &     -0&8301 &       12&97 &       12&59 &       12&35 &       11&96 &       11&63 &       11&41 &       9&84 & .&. & .&. & Galaxy\\
	SSTGLMC G059.7319+00.2144 & 59.7320 &      0&2144 &       12&95 &       12&47 &       12&23 &       11&98 &       11&71 &       11&68 &       10&23 &       7&77 & .&. & Galaxy\\
	SSTGLMC G059.9431+00.1783 & 59.9432 &      0&1783 & .&. &       15&06 &       14&19 &       13&02 &       12&40 &       11&68 &       9&76 &       6&19 & .&. & Galaxy\\
	SSTGLMC G060.9219-00.1041 & 60.9220 &     -0&1041 &       14&37 &       13&44 &       12&98 &       12&15 &       11&87 &       11&83 &       10&53 & .&. & .&. & Galaxy\\
	SSTGLMC G061.1282-00.7549 & 61.1282 &     -0&7549 &       13&60 &       12&59 &       12&32 &       12&11 &       12&14 &       11&77 &       10&63 & .&. & .&. & Galaxy\\	
	SSTGLMC G061.6453-00.7439 & 61.6454 &     -0&7439 & .&. & .&. & .&. &       12&19 &       11&91 &       11&79 &       9&53 & .&. & .&. & Galaxy\\
	SSTGLMC G060.0154+00.1111 & 60.0154 &      0&1111 & .&. & .&. & .&. &       12&88 &       10&83 &       10&01 &       9&56 & .&. &      -4&96 & Shock\\
	SSTGLMC G059.6940+00.1840 & 59.6940 &      0&1840 & .&. & .&. & .&. &       12&80 &       11&62 &       11&14 &       10&69 & .&. &      -4&82 & Shock\\
	SSTGLMC G059.7893+00.6298 & 59.7893 &      0&6298 & .&. & .&. & .&. &       12&61 &       10&75 &       10&19 &       9&27 & .&. &      -4&81 & Shock\\
	SSTGLMC G059.3780-00.2438 & 59.3781 &     -0&2438 & .&. & .&. &       14&84 &       12&76 &       11&23 &       10&63 &       10&30 &       3&14 &      -4&77 & Shock\\
	SSTGLMC G059.7973+00.0750 & 59.7973 &     0&0750 & .&. & .&. & .&. &       13&22 &       11&04 &       11&11 &       10&75 & .&. & .&. & Shock\\
	SSTGLMC G059.6366-00.1864 & 59.6366 &     -0&1864 & .&. & .&. &       14&73 &       11&48 &       9&56 &       8&79 &       8&33 &       3&13 & .&. & Shock\\
	SSTGLMC G058.6970+00.6316 & 58.6971 &      0&6316 & .&. & .&. & .&. &       12&88 &       11&41 &       10&80 &       9&70 &       3&17 &      -4&15 & Shock\\
	SSTGLMC G059.0385-00.2512 & 59.0386 &     -0&2512 & .&. &       14&47 &       13&19 &       11&32 &       10&15 &       9&89 &       8&85 &       3&67 & .&. & Shock\\
	\enddata
\tablecomments{Coordinates are in units of degrees (\degr). }
\end{deluxetable}

\begin{figure}\centering
    \includegraphics[width=0.75\textwidth,angle=0]{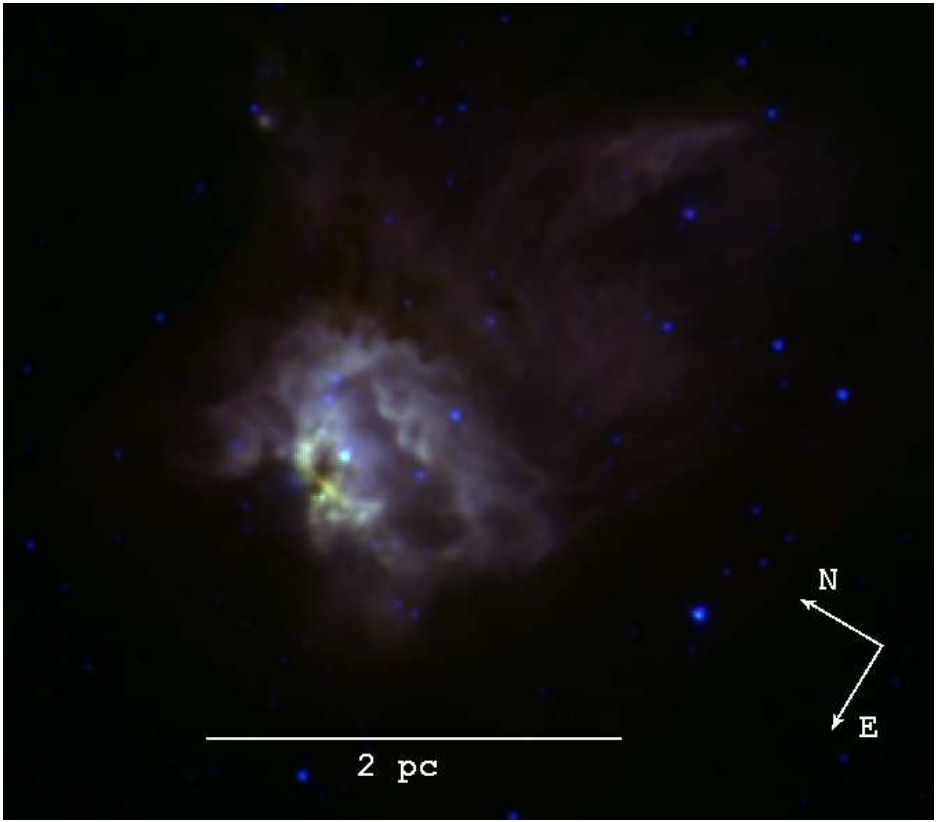}
    \caption{Composite image of the \hii region Sh88B as seen by \emph{Spitzer} (blue, green and red are [3.6], [5.8] and [8.0] respectively).}
    \label{fig:sh88b_i1i3i4}
    \end{figure}

\begin{landscape}
\begin{figure}\centering
    \includegraphics[width=1.3\textwidth,angle=0]{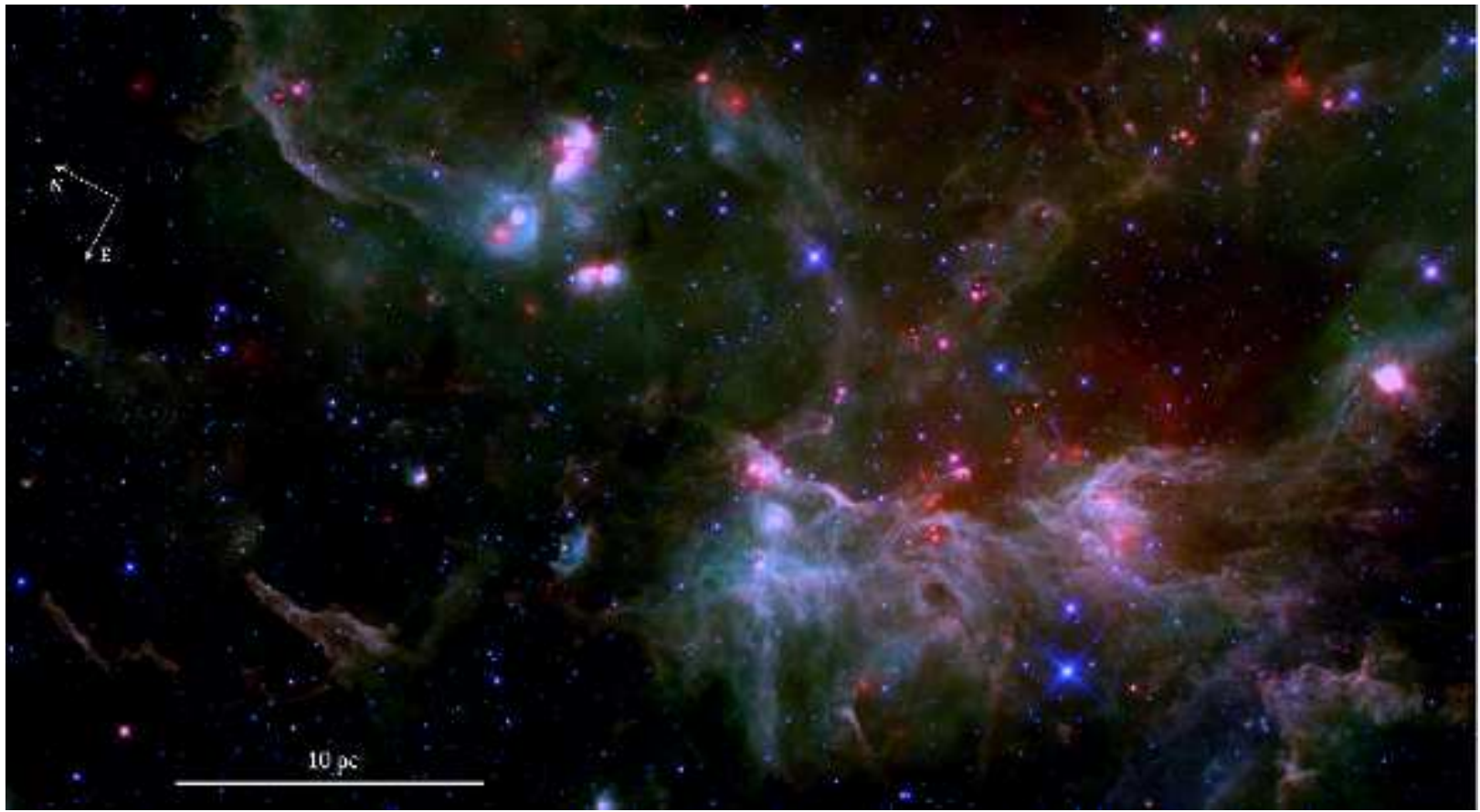}
    \caption{Composite image of the \hii region Sh86 as seen by \emph{Spitzer} (blue, green and red are [5.8], [8.0] and [24] respectively).}
    \label{fig:sh86_i3i4m1}
    \end{figure}
\end{landscape}


\clearpage


\begin{thebibliography}{}


\bibitem[Adams et al.(1987)]{adams87} Adams, F.~C., Lada, C.~J., \& Shu, F.~H., 1987, \apj, 312, 788
\bibitem[Allen et al.(2004)]{allen04} Allen, L.~E., et al., 2004, \apjs, 154, 363
\bibitem[Anantharamaiah et al.(2000)]{anan00} Anantharamaiah, K.~R., Viallefond, F., Mohan, N.~R., Goss, W.~M., \& Zhao, J.~H., 2000, \apj, 537, 613
\bibitem[Andr\'e et al.(1993)]{andre93} Andr\'e, P., Ward-Thompson, D., \& Barsony, M., 1993, \apj, 406, 122
\bibitem[Andr\'e et al.(2000)]{andre00} Andr\'e, P., Motte, F., \& Neri, R., 2000, ASP. conf., 217, 152
\bibitem[Baraffe et al.(1998)]{baraffe98} Baraffe, I., Chabrier, G., Allard, F., \& Hauschildt, P.~H., 1998, \aap, 337, 403
\bibitem[Barsony(1989)]{barsony89} Barsony, M., 1989, \apj, 345, 268
\bibitem[Benjamin et al.(2003)]{benjamin03} Benjamin, R.~A. et al., 2003, \pasp, 115, 953
\bibitem[Bertoldi(1989)]{bertoldi89} Bertoldi, F., 1989, \apj, 346, 735
\bibitem[Bica et al.(2008)]{bica08} Bica, E., Bonatto, C., \& Dutra, C.~M., 2008, \aap, 489, 1129
\bibitem[Bonnell et al.(1996)]{bonnell96} Bonnell, I.~A., Bate, M.~R., Price, N.~M., 1996, \mnras, 279, 121
\bibitem[Bowler et al.(2009)]{bowler09} Bowler, B.~P., Waller, W.~H., Megeath, S.~T., Patten, B.~M., \& Tamura, M., 2009, \aj, 137, 3685
\bibitem[Brand \& Blitz(1993)]{brand93} Brand, J. \& Blitz, L., 1993, \aap, 275, 67
\bibitem[Calvet et al.(1994)]{calvet94} Calvet, N., Hartmann, L., Kenyon, S.~J., \& Whitney, B.~A., 1994, \apj, 434, 330
\bibitem[Calvet et al.(2002)]{calvet02} Calvet, N., D'Aleesio, P., Hartmann, L., Wilner, D., Walsh, A., \& Sitko, M., 2002, \apj, 568, 1008
\bibitem[Cappa et al.(2002)]{cappa02} Cappa, C., Pineault, S., Arnal, E.~M., \& Cichowolski, S., 2002, \aap, 395, 955
\bibitem[Carey et al.(2009)]{carey09} Carey, S.~J. et al., 2009, \pasp, 121, 76
\bibitem[Carlqvist et al.(2003)]{carlqvist03} Carlqvist, P., Gahm, G.~F., Kristen, H., 2003, \aap, 403, 399
\bibitem[Chapin et al.(2008)]{chapin08} Chapin, E.~L. et al., 2008, \apj, 681, 428
\bibitem[Chavarr{\'{\i}}a et al.(2008)]{chavarria08} Chavarr{\'{\i}}a, L.~A., Allen, L.~E., Hora, J.~L., Brunt, C.~M., \& Fazio, G.~G., 2008, \apj, 682, 445
\bibitem[Cohen et al.(2003)]{cohen03} Cohen, M., Wheaton, W.~A. \& Megeath, S.~T., 2003, \aj, 126, 1090
\bibitem[D'Alessio et al.(2005)]{dalessio05} D'Alessio, P., et al., 2005, \apj, 621, 461
\bibitem[Dame et al.(2001)]{dame01} Dame, T.~M., Hartmann, D. \& Thaddeus, P., 2001, \apj, 547, 792
\bibitem[Deharveng et al.(2000)]{deharveng00} Deharveng, L., Nadeau, D., Zavagno, A., \& Caplan, J., 2000, \aap, 360, 1107
\bibitem[Dennison et al.(1998)]{dennison98} Dennison, B., Simonetti, J.~H., \& Topasna, G.~A., 1998, Publications of the Astronomical Society of Australia, 15, 147
\bibitem[Diolaiti et al.(2000)]{diolaiti00} Diolaiti, E., Bendinelli, O., Bonaccini, D., Close, L., Currie, D., \& Parmeggiani, G., 2000, \aaps, 147, 335
\bibitem[Efremov \& Elmegreen(1998)]{efremov98} Efremov, Y.~N. \& Elmegreen, B.~G., 1998, \mnras, 299, 588
\bibitem[Ehlerov\'a et al.(2001)]{ehlerova01} Ehlerov{\'a}, S., Palou{\v s}, J. \& Huchtmeier, W.~K., 2001, \aap, 374, 682
\bibitem[Elmegreen(1998)]{elmegreen98} Elmegreen, B.~G., 1998, Astronomical Society of the Pacific Conference Series, 148, 150
\bibitem[Evans et al.(2009)]{evans09} Evans, N.~J., et al., 2009, \apjs, 181, 321
\bibitem[Fazio et al.(2004)]{fazio04} Fazio, G.~G., et al., 2004, \apjs, 154, 10
\bibitem[Felli \& Harten(1981)]{felli81} Felli, M. \& Harten, R.~H., 1981, \aap, 100, 42
\bibitem[Fich \& Blitz(1984)]{fich84} Fich, M. \& Blitz, L., 1984, \apj, 279, 125
\bibitem[Flaherty(2007)]{flaherty07} Flaherty, K.~M., 2007, \apj, 663, 1069
\bibitem[Garmany \& Stencel(1992)]{garmany92} Garmany, C.~D. \& Stencel, R.~E., 1992, \aaps,94,211
\bibitem[Green(2006)]{green06} Green, D.~A., 2006, `A Catalogue of Galactic Supernova Remnants (2006 April version)', Astrophysics Group, Cavendish Laboratory, Cambridge, UK (available at "http://www.mrao.cam.ac.uk/surveys/snrs/")
\bibitem[Greene et al.(1994)]{greene94} Greene, T.~P., Wilking, B.~A., Andre, P., Young, E.~T., \& Lada, C.~J., 1994, \apj, 434, 614
\bibitem[Gritschneder et al.(2009)]{gritschneder09} Gritschneder, M., Naab, T., Walch, S., Burkert, A., \& Heitsch, F., 2009, \apjl, 694, 26
\bibitem[Guetter(1992)]{guetter92} Guetter, H.~H., 1992, \aj, 103, 197
\bibitem[Guieu et al.(2009)]{guieu09} Guieu, S. et al., 2009, \apj, 697, 787
\bibitem[Guseinov et al.(2003)]{guseinov03} Guseinov, O.~H., Ankay, A., Sezer, A., \& Tagieva, S.~O., 2003, Astronomical and Astrophysical Transactions, 22, 273
\bibitem[Gutermuth et al.(2008)]{gutermuth08} Gutermuth, R.~A. et al., 2008, \apj, 674, 336
\bibitem[Hartmann et al.(2005)]{hartmann05} Hartmann, L., Megeath, S.~T., Allen, L., Luhman, K., Calvet, N., D'Alessio, P., Franco-Hernandez, R., \&Fazio, G., 2005, \apj, 629, 681
\bibitem[Harvey et al.(2007)]{harvey07} Harvey, P., Mer{\'{\i}}n, B., Huard, T.~L., Rebull, L.~M., Chapman, N., Evans, N.~J., \& Myers, P.~C., 2007, \apj, 623, 1149
\bibitem[Hester et al.(1996)]{hester96} Hester, J.~J. et al., 1996, \aj, 111, 2349
\bibitem[Hosokawa \& Inutsuka(2006)]{hosokawa06} Hosokawa, T. \& Inutsuka, S.-i., 2006, \apjl, 648, 131
\bibitem[Hora et al.(2008)]{hora08} Hora, J.~L., et al., 2008, \aj, 135, 726
\bibitem[Hoyle et al.(2003)]{hoyle03} Hoyle, F., Shanks, T., \& Tanvir, N.~R., 2003, \mnras, 345, 269
\bibitem[Indebetouw et al.(2005)]{indebetouw05} Indebetouw, R., et al., 2005, \apj, 619, 931
\bibitem[Indebetouw et al.(2007)]{indebetouw07} Indebetouw, R., Robitaille, T.~P., Whitney, B.~A., Churchwell, E., Babler, B., Meade, M., Watson, C., \& Wolfire, M., 2007, \apj, 666, 321
\bibitem[Johnson(1958)]{johnson58} Johnson, H.~L., 1958, Lowell Observatory Bulletin, 4, 37
\bibitem[J{\o}rgensen et al.(2002)]{jorgensen02} J{\o}rgensen, J.~K., Sch{\"o}ier, F.~L. \& van Dishoeck, E.~F., 2002, \aap, 389, 908
\bibitem[Karr \& Martin(2003)]{karr03} Karr, J.~L. \& Martin, P.~G., 2003, \apj, 595, 900
\bibitem[Kharchenko et al.(2005)]{kharchenko05} Kharchenko, N.~V., Piskunov, A.~E., R{\"o}ser, S., Schilbach, E., \& Scholz, R.~D., 2005, \aap, 438, 1163
\bibitem[Koenig et al.(2008)]{koenig08} Koenig, X.~P., Allen, L.~E., Gutermuth, R.~A., Hora, J.~L., Brunt, C.~M., \& Muzerolle, J., 2008, \apj, 688, 1142
\bibitem[Kroupa(2001)]{kroupa01} Kroupa, P., 2001, \mnras, 322, 231
\bibitem[Lada(1987)]{lada87} Lada, C.~J., 1987, IAU Symposium, 115, 1
\bibitem[Lada et al.(2006)]{lada06} Lada, C.~J., et al., 2006, \aj, 131, 1574
\bibitem[Landolt-Bornstein(1982)]{LB82} Landolt-Bornstein, 1982, Bulletin of the Astronomical Institutes of Czechoslovakia, 33, V.2,subvol. b
\bibitem[Lefloch \& Lazareff(1994)]{lefloch94} Lefloch, B. \& Lazareff, B., 1994, \aap, 289, 559
\bibitem[Lortet-Zuckermann(1974)]{lortet74} Lortet-Zuckermann, M.~C., 1974, \aap, 30, 67
\bibitem[Lutz(1999)]{lutz99} Lutz, D., 1999, in The Universe as Seen by ISO, ed. P. Cox \& M.~F. Kessler, ESA~SP-427, 63
\bibitem[Makovoz et al.(2006)]{makovoz06} Makovoz, D., Khan, I., \& Masci, F., 2006, Proc. SPIE, 6065, 330
\bibitem[Marleau et al.(2008)]{marleau08} Marleau, F., et al., 2008, \aj, 136, 662
\bibitem[Massey et al.(1995)]{massey95} Massey, P., Johnson, K.~E. \& Degioia-Eastwood, K., 1995, \apj, 454, 151
\bibitem[McKee \& Ostriker(2007)]{mckee07} McKee, C.~F., \& Ostriker, E.~C., 2007, \araa, 45, 565
\bibitem[Meade et al.(2007)]{meade07} Meade, M.~R. et al., 2007, GLIMPSE~I v2.0 Data Release - http://data.spitzer.caltech.edu/popular/ glimpse/20070416\_enhanced\_v2/Documents/ glimpse1\_dataprod\_v2.0.pdf
\bibitem[Megeath et al.(2009)]{megeath09} Megeath, S.~T., Allgaier, E., Young, E., Allen, T., Pipher, J.~L., \& Wilson, T.~L., 2009, \aj, 137, 4072
\bibitem[Melioli et al.(2006)]{melioli06} Melioli, C., de Gouveia Dal Pino, E.~M., de La Reza, R. \& Raga, A., 2006, \mnras, 373, 811
\bibitem[Mer{\'{\i}}n et al.(2008)]{merin08} Mer{\'{\i}}n, B. et al., 2008, \apjs, 177, 151
\bibitem[Miao et al.(2006)]{miao06} Miao, J., White, G.~J., Nelson, R., Thompson, M., \& Morgan, L., 2006, \mnras, 369, 143
\bibitem[Mizuno et al.(2008)]{mizuno08} Mizuno, D.~R. et al., 2008, \pasp, 120, 1028
\bibitem[Mizuta et al.(2006)]{mizuta06} Mizuta, A., Kane, J.~O., Pound, M.~W., Remington, B.~A., Ryutov, D.~D., \& Takabe, H., 2006, \apj, 647, 1151
\bibitem[Molinari et al.(2008a)]{molinari08a} Molinari, S., Pezzuto, S., Cesaroni, R., Brand, J., Faustini, F. \& Testi, L., 2008, \aap, 481, 345
\bibitem[Molinari(2008b)]{molinari08} Molinari, S., 2008, 37th COSPAR Scientific Meeting, 37, 2085
\bibitem[Nomura \& Kamaya(2001)]{nomura01} Nomura, H. \& Kamaya, H., 2001, \aj, 121, 1024
\bibitem[Pascale et al.(2008)]{pascale08} Pascale, E., et al., 2008, \apj, 681, 400
\bibitem[Pigulski et al.(2000)]{pigulski00} Pigulski, A., Kolaczkowski, Z., Kopacki, G., 2000, Acta Astronomica, 50, 113
\bibitem[Price \& Bate(2009)]{price09} Price, D.~J., \& Bate, M.~R., 2009, ArXiv e-prints, 0904.4071
\bibitem[Reach et al.(2004)]{reach04} Reach, W.~T. et al., 2004, \apjs, 154, 385
\bibitem[Reach et al.(2005)]{reach05} Reach, W.~T. et al., 2005, \pasp, 117, 978
\bibitem[Reach et al.(2006)]{reach06} Reach, W.~T. et al., 2006, \aj, 131, 1479
\bibitem[Reach et al.(2009)]{reach09} Reach, W.~T. et al., 2009, \apj, 690, 683
\bibitem[Reed(2003)]{reed03} Reed, B.~C., 2003, \aj, 125, 380
\bibitem[Rieke et al.(2004)]{rieke04} Rieke, G.~H. et al., 2004, \apjs, 154, 25
\bibitem[Rieke et al.(2008)]{rieke08} Rieke, G.~H. et al., 2008, \aj, 135, 2245
\bibitem[Robitaille et al.(2007)]{robitaille07} Robitaille, T.~P., Whitney, B.~A., Indebetouw, R., \& Wood, K., 2007, \apjs, 169, 328
\bibitem[Robitaille et al.(2008)]{robitaille08} Robitaille, T.~P., et al., 2008, \aj, 136, 2413
\bibitem[Rownd \& Young(1999)]{rownd99} Rownd, B.~K. \& Young, J.~S., 1999, \aj, 118, 670
\bibitem[Schneider et al.(2006)]{schneider06} Schneider, N., Bontemps, S., Simon, R., Jakob, H., Motte, F., Miller, M., Kramer, C., \& Stutzki, J., 2006, \aap, 458, 855
\bibitem[Sharpless(1959)]{sharpless59} Sharpless, S., 1959, \apjs, 4, 257
\bibitem[Spitzer(1954)]{spitzer54} Spitzer, L.~J., 1954, \apj, 120, 1
\bibitem[Srinivasan et al.(2009)]{srinivasan09} Srinivasan, S. et al., 2009, \aj, 137, 4810
\bibitem[Stern et al.(2005)]{stern05} Stern, D. et al., 2005, \apj, 631, 163
\bibitem[Stetson(1987)]{stetson87} Stetson, P.~B., 1987, \pasp, 99, 191
\bibitem[Stil et al.(2006)]{stil06} Stil, J.~M. et al., 2006, \aj, 132, 1158
\bibitem[Stone(1970)]{stone70} Stone, M.~E., 1970, \apj, 159, 293
\bibitem[Sugitani et al.(2002)]{sugitani02} Sugitani, K. et al., 2002, \apjl, 565, 25
\bibitem[Taylor et al.(1992)]{taylor92} Taylor, A.~R., Wallace, B.~J., \& Goss, W.~M., 1992, \aj, 103, 931
\bibitem[Terebey et al.(1984)]{terebey84} Terebey, S., Shu, F.~H., \& Cassen, P., 1984, \apj, 286, 529
\bibitem[Turner(1986)]{turner86} Turner, D.~G., 1986, \aap, 167, 157
\bibitem[Urquhart et al.(2003)]{urquhart03} Urquhart, J.~S., White, G.~J., Pilbratt, G.~L., \& Fridlund, C.~V.~M., 2003, \aap, 409, 193
\bibitem[Vicente \& Alves(2005)]{vicente05} Vicente, S.~M. \& Alves, J., 2005,\aap, 441, 195
\bibitem[Wegner(2006)]{wegner06} Wegner, W., 2006, \mnras, 371, 185
\bibitem[Werner et al.(2004)]{werner04} Werner, M.~W., et al., 2004, \apjs, 154, 1
\bibitem[Xu et al.(2005)]{xu05} Xu, J.~W., Zhang, X.~Z. \& Han, J.~L., 2005,Chinese Journal of Astronomy and Astrophysics, 5, 165
\bibitem[Xue \& Wu(2008)]{xue08} Xue, R. \& Wu, Y., 2008,\apj, 680, 446
\bibitem[Zavagno et al.(2006)]{zavagno06} Zavagno, A., Deharveng, L, Comer{\'o}n, F., Bran, J., Massi, F., Caplan, J., \& Russeil, D., 2006, \aap, 446, 171


\end{thebibliography}
\end{document}